\documentclass[12pt]{article}

\usepackage{amsmath}
\usepackage{amssymb}
\usepackage{mathrsfs}
\usepackage[T1]{fontenc}
\usepackage{amsthm}
\usepackage{enumerate}
\def\eqn{equation}

\def\tfn{transformation}

\def\sm{sigma model}
\def\pl{Poisson--Lie}

\def\dd{Drinfeld double}

\def\4diml{four-dimensional}
\def\-1{^{-1}}
\def\half{\frac{1}{2}}
\def\coor{coordinate}
\def\real{\mathbb{R}}

\def \unit {{\bf 1}}

\def\e{{\rm e}}

\def\wt{\tilde}
\def\wwt{\widetilde}
\def\sm{sigma model}

\def\pltd{Poisson--Lie T-dualit}

\begin{document}

%\frontmatter

%%%%%%%%%%%%%%%%%%%%%%%%%%%%%%%%%%%%%%%%%%%%%%%%%%%%%%%%%%%%%%%%%%%%%%%%%%%%%%%
%% Titlepage
%%%%%%%%%%%%%%%%%%%%%%%%%%%%%%%%%%%%%%%%%%%%%%%%%%%%%%%%%%%%%%%%%%%%%%%%%%%%%%%

\title{Plane-parallel waves as
duals of the flat background}
\author{Ladislav Hlavat\'y\footnote{hlavaty@fjfi.cvut.cz}, Ivo Petr\footnote{ivo.petr@fjfi.cvut.cz, also at Faculty of Information Technology,
Czech Technical University in Prague,
Th\' akurova 9, Prague 6, 160 00,
Czech Republic}
}
\maketitle
%\address
{Faculty of Nuclear Sciences and Physical Engineering, Czech Technical University in Prague,  B\v rehov\'a 7, 115 19 Prague 1, Czech Republic}

\begin{abstract}
{We give a classification of non-Abelian T-duals of the flat metric in $D=4$ dimensions with respect to
the four-dimensional continuous subgroups of the Poincaré group.} After dualizing
the flat background, we identify majority of dual models as conformal sigma models in
plane-parallel wave backgrounds, most of them having torsion. We give their form in
Brinkmann coordinates. We find, besides the plane-parallel waves, several diagonalizable curved metrics with nontrivial scalar curvature and torsion. Using the non-Abelian T-duality, we find general solution of the classical field equations for
all the sigma models in terms of d'Alembert solutions of the wave \eqn{}.
\end{abstract}

%\keywords{Sigma Models; Penrose limit and pp-wave background; String Duality.}

%\ccode{PACS numbers: 11.10.Lm, 11.27.+d,  04.60.Cf, 02.30.Ik}

\tableofcontents

%%%%%%%%%%%%%%%%%%%%%%%%%%%%%%%%%%%%%%%%%%%%%%%%%%%%%%%%%%%%%%%%%%%%%%%%%%%%%%%
%% Introduction
%%%%%%%%%%%%%%%%%%%%%%%%%%%%%%%%%%%%%%%%%%%%%%%%%%%%%%%%%%%%%%%%%%%%%%%%%%%%%%%

\section{Introduction}\label{intro}
String theory in curved and/or time-dependent background can be formulated as a sigma model
satisfying supplementary conditions. Finding solutions of equations of motion in such
backgrounds is usually very complicated. That is why every solvable case attracts
considerable attention. An example of such a model is a string theory in the homogeneous
plane-parallel wave background, solved in Ref. \cite{papa} in terms of Bessel functions.
Plane-parallel (pp-)wave backgrounds in string theory have been repeatedly investigated in the
past (see e.g. references in \cite{tsey:revExSol}). They not only give solvable models
\cite{BLPT}, but also allow one to study the behavior of strings near spacetime singularities
\cite{HorSteif,VegaSan}. Moreover, to extract information about string behavior in a general
curved background, one can take the Penrose limit \cite{Penrose}, extended to fields in
string theory in \cite{Gueven}, and study string behavior in the resulting plane-wave
background.

{Particular cases of four-dimensional pp-wave background in Rosen coordinates obtained from
gauged WZW (Wess--Zumino--Witten) models were given in \cite{Sfetseyt94},\cite{tseyt94}, and \cite{tsey:revExSol}, as
\begin{equation}\label{sfets} ds^2= dudv + {g_1 (u') \over g_1 (u')g_2 (u)  + q^2 }\ dx_1^2 +
 {g_2 (u)\over g_1 (u')g_2(u)  + q^2} \ dx_2^2,
\end{equation}
$$B_{12 }= {q \over  g_1 (u') g_2 (u) + q^2},$$ where $u'=
au + d \ $ ($a,d=const$) and the functions $g_i$ can take any pair of the following values:
\begin{equation}\label{g1g2}
g(u) = 1\ , \ \ u^2\  ,  \ \  \tanh^2 u \ , \ \  \tan^2 u \ , \ \
 u^{-2}\ , \ \ \coth^2 u\ , \ \ \cot^2 u.
\end{equation}
It is mentioned in \cite{Sfetseyt94} that this background is dual to the flat space for $
g_1=1,\ g_2=u^2$. We shall show that several other cases of these backgrounds are dual to
the flat space as well. Moreover, we shall use this fact for finding general solutions of
classical \sm{} field \eqn s in these pp-wave backgrounds.} We fing, beside pp-waves, several curved backgrounds with diagonalizable metrics resembling black hole
\cite{horava:blackhole} and cosmological \cite{nappiw92} solutions {and we check that solutions obtained by non-Abelian T-duality satisfy sigma model field equations in these backgrounds as well.}

We understand the non-Abelian T-duality \cite{delaossa:1992vc} as a special case of \pltd y
\cite{klise} {based on the structure of the \dd.} For technical reasons we shall restrict to
four spacetime dimensions, but the discussion can be  extended to higher dimensions using
the spectator fields or subgroups of Poincaré group in higher dimensions. Investigation of
conformal invariance of pp-waves in higher dimensions can be found e.g. in
\cite{dhh:vca},\cite{dhh:spfgw}.

The plan of the paper is the following. In the {next two
sections we describe the method whereby the \pltd y is used as a tool
for the construction of dual models and their solution.
In section \ref{ppwaves},  we review relevant properties concerning strings
in the pp-wave background. Detailed discussion of particular examples is given
in section \ref{examples}. Section \ref{results} summarizes results of dualization
with respect to various subgroups of the Poincaré group.
Subalgebras corresponding to these subgroups are listed in the appendix}.% \ref{appA}.}

%%%%%%%%%%%%%%%%%%%%%%%%%%%%%%%%%%%%%%%%%%%%%%%%%%%%%%%%%%%%%%%%%%%%%%%%%%%%%%%
%% Main matter
%%%%%%%%%%%%%%%%%%%%%%%%%%%%%%%%%%%%%%%%%%%%%%%%%%%%%%%%%%%%%%%%%%%%%%%%%%%%%%%

%\mainmatter
\section{Non-Abelian T-duality}\label{secPLT}

{The sigma model on a manifold $M$ is given by the classical action
\begin{align}
\label{sma}
        &S_{F}[X]=
-\int d\sigma_+d\sigma_-\,(\partial_{-}X^{\mu}F_{\mu\nu}(X)\partial_{+}X^{\nu})=\\
=\nonumber \half \int d\tau
d\sigma\,&\left[-\partial_{\tau}X^{\mu}G_{\mu\nu}(X)\partial_{\tau}X^{\nu}+
\partial_{\sigma}X^{\mu}G_{\mu\nu}(X)\partial_{\sigma}X^{\nu}
-2\partial_{\tau}X^{\mu}B_{\mu\nu}(X)\partial_{\sigma}X^{\nu}\right],
\end{align}
where  $F$ is a second order tensor field on $M$, with the metric
and the NS--NS 2-form (torsion potential; NS standing for Neveu--Schwarz) given by the symmetric and
antisymmetric part of $F$:
    \[ G_{\mu\nu}=\half(F_{\mu\nu}+F_{\nu\mu}),\
    B_{\mu\nu}=\half(F_{\mu\nu}-F_{\nu\mu}).
    \]
    The worldsheet coordinates are
    \[
   \sigma_+=\frac{1}{\sqrt{2}}(\tau+\sigma),\ \sigma_-  =\frac{1}{\sqrt{2}}(\tau-\sigma).
    \]
    The functions
$X^{\mu}$ %$$:\mathbb{{R}}^{2}\rightarrow\mathbb{R}$, $\mu =1,2,\ldots,dim(G),$
are determined by the composition $X^{\mu}(\tau,\sigma)=x^{\mu}(
X(\tau,\sigma))$, where $X:{\mathbb{R}^{2}}\ni(\tau,\sigma)\mapsto
X(\tau,\sigma)\in M$ and $x^{\mu}:\mathbf{U}_{p}\rightarrow
\mathbb{R}$ are components of a coordinate map on a neighborhood
$\mathbf{U}_{p}$ of an element $X(\tau,\sigma)=p\in M$.

The non-Abelian T-duality}
\cite{delaossa:1992vc} of \sm s is a special case of {\pltd y
\cite{klise},\cite{klim:proc} that can be formulated} by virtue of
the \dd {} -- a connected Lie group whose Lie algebra
${\mathfrak{d}}$ can be decomposed into a pair of equally
dimensional subalgebras ${\mathfrak{g}},\tilde{\mathfrak g}$ {that
are} maximally isotropic with respect to a symmetric ad-invariant
non-degenerate bilinear form $\langle.,.\rangle$ on ${\mathfrak{d}}$.

The \dd{} {suitable} for non-Abelian T-duality is the semidirect
product $G\ltimes\wwt G$, where
the group $G$ %suitable for a dualizable sigma model
can be taken as a subgroup of the isometry group of the {background, which, in our case, will be
flat}. The group $\wwt G$ {has to} be chosen Abelian in order to satisfy the conditions of
dualizability of the \sm\, \cite{klise}. {We shall focus on the case, when the isometry subgroup acts
freely and transitively on the manifold, {so that we can make the identification $G\approx M$. This is}
usually referred to as atomic duality. Let us summarize the main points of the construction
of dual models.

Given the four-dimensional subgroup $G$ of the isometry group
generated by Killing vectors of the flat metric, the tensor $F$ is
symmetric and can be written as
    \begin{equation}\label{met}
        F_{\mu\nu}(x)=G_{\mu\nu}(x)=e_{\mu}^{a}(g(x))(E_0)_{ab}e_{\nu}^{b}(g(x)),
    \end{equation}where $E_0$ is a constant {non-singular} symmetric matrix, and  $e_{\mu}^{a}(g(x))$ are components of the right-invariant forms
$(dg)g^{-1}$ %)_{\mu}^{a}$
expressed in coordinates $\{x^\mu\}$ on the group $G$ and the basis of its Lie algebra
$\{T_a\}$.

{Denoting the mutually dual bases of ${\mathfrak{g}}$ and Abelian
$\tilde{\mathfrak g}$ as $\{T_{i}\}$, $\{{\tilde T}^{j}\}$, we
construct subspaces $\varepsilon^{+}=Span(T_{i}+E_{0,ij}{\tilde
T}^{j})$, $\varepsilon^{-}=Span(T_{i}-E_{0,ji}{\tilde T}^{j})$ that
are orthogonal w.r.t. $\langle,\rangle$ and span the whole Lie algebra
$\mathfrak d$. The field equations for the \sm{} on the group $G$
can be rewritten as the \eqn
    \begin{equation}\label{deqn}
        <(\partial_{\pm}l)l^{-1},\varepsilon^{\pm}>=0,
    \end{equation}
for mapping $l$ from the worldsheet in $\mathbb{R}^2$ into the Drinfeld double $D$.}

Due to Drinfeld, there exists a unique decomposition (at least in the vicinity of the unit
element of $D$) of an arbitrary element $l$ of $D$ as a product of elements
%$l=g\, \tilde h$
from $G$ and $\tilde G$. Solutions of \eqn{} (\ref{deqn}) and solution of the \eqn s of
motion for the \sm {} $X^\mu(\tau,\sigma) = x^\mu( g(\tau,\sigma))$ are related by
\begin{equation*}        l(\tau,\sigma)=g(\tau,\sigma){\tilde{h}}(\tau,\sigma)\in D,
    \end{equation*}
where ${\tilde {h}}(\tau,\sigma)\in {\tilde{G}}$ fulfills the equations
\begin{equation}\partial_\tau
\tilde
h_j = %-A_{\tau,j}:=
-v^\lambda_jG_{\lambda\nu}\partial_\sigma X^\nu \label{btp},\end{equation}
\begin{equation}\partial_\sigma \tilde h_j =
%-A_{\sigma,j}:=
-v^\lambda_jG_{\lambda\nu}\partial_\tau X^\nu,
 \label{btm}
 \end{equation}
{with $v_j^\lambda$ representing components} of the left-invariant
fields $v_j$ on $G$ in the group coordinates $x^\mu$.
% and ${\tilde {h}}_j(\tau,\sigma)$ are coordinates of ${\tilde {h}}(\tau,\sigma)$ in .

The metric and the torsion potential of the non-Abelian T-dual model can be obtained from the
tensor $\wwt F$:
\begin{equation}\label{tildeF} \wwt F_{\mu\nu}(\wt x)=[E_{0}+\wwt\Pi(\wt
g(\wt x))]^{-1},
\end{equation}where the matrix $\wwt \Pi$ is given by the adjoint representation of the Abelian subgroup
$\wwt G$ on the Lie algebra of the Drinfeld double in the mutually dual bases
\begin{equation*}\label{wwtPi}
Ad(\wt g)^{T}=\left(
                \begin{array}{cc}
                    \unit & 0 \\
                   \wwt\Pi(\wt g) & \unit \\
                \end{array}
      \right).
\end{equation*}

The relation between the solution $X^\mu(\tau,\sigma) $ of the \eqn s of motion of the sigma model
given by $F$ {and the solution} $\wt X^\mu(\tau,\sigma):=\wt x^\mu (\wt g(\tau,\sigma))$ of the
\sm{} given by $\wwt F$ follows from two possible decompositions of elements $l$ of the
Drinfeld double:
\begin{equation}\label{lghtgth}
         g(\tau,\sigma)\tilde h(\tau,\sigma)=\tilde g(\tau,\sigma)h(\tau,\sigma)
   , \end{equation}where $g, h \in G,\ \wt g, \wt h\in \wt G$. The map $ \wt h:\real^2\rightarrow \tilde G$ that we need
for
this transformation is obtained from equations  %\cite{kli:pltd}
\eqref{btp},\eqref{btm}.

\section{Solving the classical \sm{} \eqn s by non-Abelian T-duality}\label{solvePLT}
Equation (\ref{lghtgth}) defines the \pl {} transformation
between solutions of the \eqn s of motion {of the original \sm {} and
its dual. Its} application may be rather complicated. To use {it for
finding the solution of the dual} model, the following three steps must
be achieved:
\begin{itemize}
\item Step 1: One has to know the solution $X^\mu(\tau,\sigma)$ of the \sm{} given by $F_{\mu\nu}(x)$.
\item Step 2: {Given} $X^\mu(\tau,\sigma)$, one has to find $\wt h(\tau,\sigma)$,
i.e. solve the system of PDEs \eqref{btp},\eqref{btm}.
\item Step 3: {Given $l(\tau,\sigma)= g(\tau,\sigma)\tilde h(\tau,\sigma)\in D$, one has to find the dual decomposition
$l(\tau,\sigma)=\tilde g(\tau,\sigma)h(\tau,\sigma)$, where $\wt g(\tau,\sigma)\in \wt G$,
$h(\tau,\sigma)\in G$}. Functions $\wt X^\mu(\tau,\sigma):=\tilde x^\mu(\tilde g(\tau,\sigma))$ then solve
the field \eqn s of the dual \sm.
\end{itemize}

For simplicity, we will restrict consideration to four spacetime
dimensions. Our convention for the flat metric in the spacetime coordinates $(t,x,y,z)$ is
\begin{equation*}
\eta=\text{diag}(-1,1,1,1).
\end{equation*}
{It is easy to find solutions for the equations following from the
flat metric in coordinates $x^I\in\{t,x,y,z\}$, as they reduce to
two-dimensional wave equations
\begin{equation}\label{wavesolns}
\partial_{\tau}^2W^J- \partial_{\sigma}^2W^J=0,\ \quad J=t,x,y,z.
\end{equation}
However, we need to identify the group $G$ with the manifold, i.e. find appropriate coordinate transformation between $(t,x,y,z)$ and the coordinates parametrizing the group. Choosing the parametrization of group elements as
\begin{equation}\label{gparametrization}
    g=g(x^\mu)=e^{x^{1}T_{1}}e^{x^{2}T_{2}}e^{x^{3}T_{3}}e^{x^{4}T_{4}},
\end{equation}
where $T_j$ form the basis of the Lie algebra of the group, we may calculate the algebra of left-invariant fields
$$v_j=v_j^\mu\frac{\partial}{\partial x^\mu},\quad j=1,\ldots,4,$$
and compare it with the chosen four-dimensional subalgebra of Killing vectors $(\mathcal{K}_i)$ of the flat metric in coordinates $(t,x,y,z)$.  The comparison then may give the coordinate \tfn{}
$$ x^\mu=x^\mu(t,x,y,z)$$ as a solution to a set of PDEs.
}

The right-hand sides of the PDEs \eqref{btp},\eqref{btm}, solved in
step 2, are invariant w.r.t coordinate transformation. This means
that we can express the right-hand sides in terms of the coordinates
$(t,x,y,z)$ and  use the { Killing fields $\mathcal{K}_j$ %\eqref{Poincare killings}
instead} of the left-invariant fields on $G$. The \eqn s
\eqref{btp} and \eqref{btm} then acquire the form
\begin{equation}\partial_\tau
\tilde h_j = -\mathcal{K}^I_j\eta_{IJ}\partial_\sigma W^J \label{btp2},\end{equation}
\begin{equation}\partial_\sigma \tilde h_j =
-\mathcal{K}^I_j\eta_{IJ}\partial_\tau W^J,
 \label{btm2}
 \end{equation}
where the $W^J$ are solutions of two-dimensional wave \eqn s
\eqref{wavesolns}, {and $\mathcal{K}_j^I$ are the components of
Killing vectors in coordinates $(t,x,y,z)$.}

Step 3 represents in general rather complicated problem related to the
Baker--Campbell--Hausdorff formula. Its solution simplifies substantially when the adjoint
representation of the Lie algebra $\mathfrak g$ is faithful. Let
\begin{equation}\label{rozklad dd}
    l=g\tilde h=\tilde g h,\ \ \ g,h\in G,\ \ \tilde g, \tilde h \in \tilde G=
        \real^4,
\end{equation}
and assume that the parametrizations of $g,h,\tilde g, \tilde h $ are
\begin{eqnarray*}\label{ght}
  g=e^{x^{1}T_{1}}e^{x^{2}T_{2}}e^{x^{3}T_{3}}e^{x^{4}T_{4}},&&\  \wt h=e^{\wt h_{1}\wt T^{1}}e^{\wt h_{2}\wt T^{2}}e^{\wt h_{3}\wt T^{3}}e^{\wt h_{4}\wt T^{4}}
          ,\\ \label{gth}
 \wt g=e^{\wt x_{1}\wt T
 ^{1}}e^{\wt x_{2}\wt T^{2}}e^{\wt x_{3}\wt T^{3}}e^{\wt x_{4}\wt T^{4}}, && h=e^{h^{1} T
 _{1}}e^{h^{2}T_{2}}e^{h^{3} T_{3}}e^{h^{4} T_{4}}.
\end{eqnarray*}

The variables $x^j,\wt h_k$ and $\wt x_j, h^k$ represent two sets of coordinates in
(the vicinity of the unit of) the \dd. To express $\wt x_j, h^k$ in terms of $x^j,\wt h_k$, we can
use a representation $r$ {of an element of} the semi-Abelian \dd{} in the form of block
matrices $(\dim {\mathfrak g}+1)\times (\dim {\mathfrak g}+1)$, such that
\begin{equation*}\label{Drep}
   r(g)= \left(
\begin{array}{cc}
Ad\, g & 0 \\
 0& 1
\end{array}
\right),\ \ \   r(\wt h)= \left(
\begin{array}{cc}
{\bf 1}& 0 \\
 v(\wt h)& 1
\end{array}
\right),
\end{equation*}
where $v(\wt h)=(\wt h_1,\ldots,\wt h_{\dim\mathfrak g})$.
From the equation (\ref{rozklad dd}) we then get
\begin{equation}\label{replgh}
   r(l)=r(g\wt h)= \left(
\begin{array}{cc}
Ad\, g & 0 \\
 v(\wt h)& 1
\end{array}
\right) =r(\wt g  h)= \left(
\begin{array}{cc}
Ad\, h & 0 \\
 v(\wt g)\cdot(Ad\, h )& 1
\end{array}
\right). \end{equation} If the adjoint representation of the Lie algebra $\mathfrak g$ is
faithful, then the representation $r$ of the \dd {} is faithful as well, and the relation
(\ref{replgh}) gives a system of \eqn s for $\tilde x_j$ and $h^j$. If not, we can try to
use formula \begin{equation}\label{BHC}
    \e^{A}\e^{B}=\e^{\exp(adA) B}\e^A
\end{equation}
to permute the elements of $G$ and $\wt G$ in (\ref{rozklad dd}) and
express the coordinates $\wt x_j, h^k$ in terms of $x^j,\wt h_k$.

In the {following sections we shall apply the above given three steps of} the \pl {}
transformation to solve the \sm{} field equations in curved backgrounds dual to the flat
metric.
\section{Strings in the pp-wave background}\label{ppwaves}

We will be interested in the special subclass of metrics called
pp-waves. Their metric in the so called Brinkmann coordinates
$(u,v,z_3,z_4,\ldots,z_D)$ can be written as
\begin{equation}\label{BLPT_pp_wave}
  ds^{2}= 2dudv-K(u,\vec z)du^{2}+  d\vec{z}{}\,^{2},
\end{equation}
where $d\vec{z}{}\,^{2}$ is the Euclidean metric in the transversal space with coordinates
$\vec{z}=(z_3,z_4,\ldots,z_D)$. We denote the number of transversal coordinates by $d$,
such that $D=2+d$. The NS--NS 2-form of particular interest to us has the form
\begin{equation}\label{BLPT_B_and_dil}
B = B_{j}(u,\vec z)du\wedge dz_j.
\end{equation}
The metric \eqref{BLPT_pp_wave} has covariantly constant null
Killing vector $\partial _v$ and particularly simple curvature
properties, because the Ricci tensor has only one nonzero component
$$R_{uu}=\half(\partial_3^2K+\partial_4^2K+\ldots+\partial_D^2K),$$
and the scalar curvature vanishes. The one-loop conformal
invariance conditions for the \sm
\begin{eqnarray}
\label{bt1} 0 & = &
R_{\mu\nu}-\bigtriangledown_\mu\bigtriangledown_\nu\Phi-
\frac{1}{4}H_{\mu\kappa\lambda}{H_\nu}^{\kappa\lambda},
\\
 \label{bt2} 0 & = & \bigtriangledown^\mu\Phi H_{\mu\kappa\lambda}+\bigtriangledown^\mu H_{\mu\kappa\lambda}\,,
\\
\label{bt3} 0 & = & R-2\bigtriangledown_\mu\bigtriangledown^\mu\Phi-
\bigtriangledown_\mu\Phi\bigtriangledown^\mu\Phi-
\frac{1}{12}H_{\mu\kappa\lambda}H^{\mu\kappa\lambda},
\end{eqnarray}
where $H=dB$, can be solved in some special cases. {One of them is that of
the model in the background resulting from the Penrose--{G{\" u}ven}
limit \cite{Penrose,Gueven}, with}
\begin{equation}\label{Penrose_K}
K(u,\vec z)=K_{ij}(u)z_iz_j,
\end{equation}
and the torsion
\begin{equation}\label{Gueven_torsion}
    H=H_{ij}(u)du\wedge dz_i\wedge dz_j
\end{equation}
that follows from the NS--NS 2-form
\eqref{BLPT_B_and_dil} if $B_j(u,\vec{z})$ is linear in $z$.
The one-loop conformal invariance conditions  then simplify to
solvable differential equation for the dilaton $\Phi=\Phi(u)$:
\begin{equation}\label{dilaton_eqn}
   \Phi''(u)-K_{jj}(u)+\frac{1}{4} H_{ij}(u)H_{ij}(u)=0.
\end{equation}

We are going to show that {the sigma models in pp-wave backgrounds} with special forms of the
functions $K_{ij}$ \eqref{Penrose_K} and $H_{ij}$ \eqref{Gueven_torsion} can be obtained as
non-Abelian T-duals of {\sm s in} the flat background.
The Killing vectors of the flat metric $\eta=\text{diag}(-1,1,1,1)$ in coordinates $(t,x,y,z)$ are
\begin{equation}\label{Poincare killings}
    P_0=\partial_t,\ P_j=\partial_j,\ L_j=-\varepsilon_{ijk}x^j\partial_k,\
    M_j=-x^j\partial_t-t\partial_j,
\end{equation}and form the ten-dimensional Poincaré Lie algebra.
To apply the (atomic) non-Abelian T-duality on \sm s in the flat {background,} we shall
need four-dimensional subalgebras of the Poincaré Lie algebra, classified in \cite{PWZ}.

For $K(u,\vec{z})$ in \eqref{BLPT_pp_wave} at most quadratic in
transversal coordinates, one can find \tfn s that bring it to the form
\eqref{Penrose_K}. {In the following, we will be able to bring the
metrics of the resulting dual models with vanishing scalar curvature
to the form \eqref{BLPT_pp_wave}, where}
\begin{equation}\label{Penrose_KK}
K(u,\vec z)=K_3(u)z_3^2+K_4(u)z_4^2,
\end{equation}
and the torsion is
$$ H=H(u)\,du\wedge dz_3\wedge dz_4.$$
Classical field equations of sigma model \eqref{sma} in such a
background then read
\begin{equation}\label{pohrceu}
    \partial^2_{\tau}U-  \partial^2_{\sigma}U=0,
\end{equation}
\begin{equation}\label{pohrcez3}
    \partial^2_{\tau}Z_3-\partial^2_{\sigma}Z_3=K_3(U)\left[(\partial_{\sigma}U)^2-(\partial_{\tau}U)^2\right]Z_3
    -H(U)\left[\partial_{\sigma}Z_4 \partial_{\tau}U-\partial_{\tau}Z_4\partial_{\sigma}U\right],
\end{equation}
\begin{equation}\label{pohrcez4}
    \partial^2_{\tau}Z_4-\partial^2_{\sigma}Z_4=K_4(U)\left[(\partial_{\sigma}U)^2-(\partial_{\tau}U)^2\right]Z_4
    +H(U)\left[\partial_{\sigma}Z_3 \partial_{\tau}U-\partial_{\tau}Z_3\partial_{\sigma}U\right],
\end{equation}
\begin{eqnarray}\nonumber
    \partial^2_{\tau}V- \partial^2_{\sigma}V&=&
         H(U)\left[\partial_{\sigma}Z_4\partial_{\tau}Z_3-\partial_{\sigma}Z_3\partial_{\tau}Z_4 \right]+
        \\&&  \sum_{j=3}^4\Big\{2K_j(U)Z_j\left[\partial_{\tau}Z_j\partial_{\tau}U-\partial_{\sigma}Z_j \partial_{\sigma}U\right]+ \label{pohrcev}
        \\&&(Z_j)^2\left[\half K_j'(U)\left[(\partial_{\tau}U)^2-(\partial_{\sigma}U)^2\right]
    +K_j(U)\left(\partial^2_{\tau}U- \partial^2_{\sigma}U\right)\right]\Big\}. \nonumber
\end{eqnarray}
For string backgrounds, the last equation can be replaced by the so-called string conditions for $X^\mu=(U,V,Z_3,Z_4)$
\begin{equation}\label{stringcond1}
    \partial_\tau X^\mu G_{\mu\nu}(X)\partial_\tau X^\nu+\partial_\sigma X^\mu
    G_{\mu\nu}(X)\partial_\sigma X^\nu=0,
\end{equation}
\begin{equation}\label{stringcond2}
 \partial_\tau X^\mu G_{\mu\nu}(X)\partial_\sigma X^\nu=0.
\end{equation}
Conditions \eqref{stringcond1},\eqref{stringcond2} for the pp-wave
with function $K$ given by \eqref{Penrose_KK}
 yield
\begin{equation*}\label{stringcond1pp}
    2\partial_{\tau}U \partial_{\tau}V+\sum_{j=3}^4\Big\{(\partial_{\tau}Z_j)^2-(\partial_{\tau}U)^2 K_j(U)(Z_j)^2
     \Big\}+(\tau\rightarrow\sigma)=0,
\end{equation*}
\begin{equation*}\label{stringcond2pp}
    \partial_{\tau}U \partial_{\sigma}V+\partial_{\tau}V \partial_{\sigma}U +\sum_{j=3}^4\Big\{\partial_{\tau}Z_j\partial_{\sigma}Z_j-\partial_{\tau}U \partial_{\sigma}U K_j(U)(Z_j)^2
     \Big\}=0.
\end{equation*}
Compatibility of these two first order equations for
$V=V(\tau,\sigma)$ is guaranteed by the equations
\eqref{pohrceu} -- %\ref{pohrcez3}
\eqref{pohrcez4}.

Note that for nonvanishing torsion, both $Z_3$ and $Z_4$ appear in
(\ref{pohrcez3},\ref{pohrcez4}), so that even in the light-cone gauge $U=\kappa \tau$
these equations do not separate, and it can be rather difficult to solve them in the usual way
using Fourier mode expansion. Nevertheless, the T-duality gives a {method to obtain the
general solution}.

\section{Examples}\label{examples}
\subsection{Example 1 -- subalgebra $S_{27}$}\label{example} We shall
illustrate the above described methods of non-Abelian dualization of
the flat metric on the example of Killing vectors
\begin{eqnarray} \nonumber
    \mathcal{K}_1&=&M_3=-z\partial_t-t\partial_z,
    \\ \label{killings27} \mathcal{K}_2&=&L_2+M_1=-x\partial_t-(t+z)\partial_x+x\partial_z,
    \\ \nonumber \mathcal{K}_3&=&L_1 -
    M_2=y\partial_t+(t+z)\partial_y,-y\partial_z,
    \\ \nonumber \mathcal{K}_4&=&P_0-P_3=\partial_t-\partial_z
\end{eqnarray}
that span subalgebra $S_{27}$ (see the appendix). Their nonvanishing commutation relations
are
\begin{equation}\label{comrel27}
    [\mathcal{K}_1,\mathcal{K}_2]=-\mathcal{K}_2,\qquad [\mathcal{K}_1,\mathcal{K}_3]=-\mathcal{K}_3,\qquad [\mathcal{K}_1,\mathcal{K}_4]=-\mathcal{K}_4.
\end{equation}
\subsubsection{Duals to the flat metric}
Using the parametrization \eqref{gparametrization} of the isometry
subgroup $G$, where $T_\mu$ are elements of its Lie algebra
commuting as in %with commutation relations
\eqref{comrel27}, we get
the basis of left-invariant fields on $G$
\begin{equation*}
\label{leftverfields27}
  v_1 = \partial_1+x^2\partial_2+x^3\partial_3+x^4\partial_4 ,
\end{equation*}
\begin{equation*}
v_2 = \partial_2 ,\qquad v_3= \partial_3 ,\qquad   v_4 =
\partial_4.
\end{equation*}
Identifying the Killing vectors \eqref{killings27} with these
left-invariant fields, we get the transformation of coordinates on the
flat manifold
\begin{align}
\label{txyz27}  t  &= \frac{1}{2} e^{-x^1} \left((x^2)^2+(x^3)^2+1\right)+x^4, &  x &= -e^{-x^1} x^2,\\
\nonumber z &=-\frac{1}{2} e^{-x^1} \left((x^2)^2+(x^3)^2-1\right)-x^4, & y &= e^{-x^1} x^3
\end{align}
that gives the flat metric in the group coordinates $x^\mu$
\begin{equation}\label{mtk27}
    G_{\mu\nu}(x)=F_{\mu\nu}(x)=\left(
\begin{array}{cccc}
 0 & 0 & 0 & e^{-x^1} \\
 0 & e^{-2 x^1} & 0 & 0 \\
 0 & 0 & e^{-2 x^1} & 0 \\
 e^{-x^1} & 0 & 0 & 0
\end{array}
\right).
\end{equation}
{This can be obtained from} the formula \eqref{met} if one chooses}
\begin{equation*}\label{E027}
    E_0=\left(
\begin{array}{cccc}
 0 & 0 & 0 & 1 \\
 0 & 1 & 0 & 0 \\
 0 & 0 & 1 & 0 \\
 1 & 0 & 0 & 0
\end{array}
\right).
\end{equation*}

The dual tensor $\wwt F$  can be then found from the formula \eqref{tildeF} as
\begin{equation*}\label{tildemtz27}
    \wwt F_{\mu\nu}(\wt x)=\left(
\begin{array}{cccc}
 0 & 0 & 0 & \frac{1}{1-\wt x_4} \\
 0 & 1 & 0 & \frac{\wt x^2}{1-\wt x_4} \\
 0 & 0 & 1 & \frac{\wt x_3}{1-\wt x_4} \\
 \frac{1}{\wt x_4+1} & -\frac{\wt x_2}{\wt x_4+1} & -\frac{\wt x_3}{\wt x_4+1} &
   \frac{\wt x_2^2+\wt x_3^2}{\wt x_4^2-1}
\end{array}
\right).
\end{equation*}
Scalar curvature corresponding to the metric obtained from the symmetric part of this tensor %of \eqref{tildemtz27}
vanishes, and Ricci tensor has only one nonvanishing component
$$ \wwt R_{44}=-\frac{4}{({\wt x_4}^2-1)}.$$
This suggests that dual metric could be of the pp-wave form. Indeed, the transformation of
coordinates for $|\wt x_4|>1$
\begin{align}
\label{xtobrink27} \wt x_1 &= v-\frac{1}{2} ({z_3}^2+{z_4}^2) \coth ({u}),& \wt x_2 &= z_3,\\ \nonumber \wt x_4 &=\coth ({u}),  & \wt x_3 &= z_4,
\end{align}
brings the components of the tensor $\wwt F$ into the form\begin{equation}\label{dual27B}
\wt F=   \left(
\begin{array}{cccc}
 2\,\frac{{z_3}^2+{z_4}^2}{ {\sinh}^2({u})} & 1-\coth (u) & \frac{{z_3}}{ {\sinh}^2({u})}  & \frac{{z_4}}{ {\sinh}^2({u})}  \\
 1+\coth (u) & 0 & 0 & 0 \\
 -\frac{{z_3}}{ {\sinh}^2({u})} & 0 & 1 & 0 \\
 -\frac{{z_4}}{ {\sinh}^2({u})} & 0 & 0 & 1
\end{array}
\right).
\end{equation}
The symmetric part %of this tensor
yields pp-wave metric in the Brinkmann form
\begin{equation}\label{dualmtkBrink27}
ds^{2}= 2dudv+2\,\frac{{z_3}^2+{z_4}^2}{
{\sinh}^2({u})}du^{2}+d{z_3}^2+d{z_4}^2.\end{equation} Torsion obtained from the
antisymmetric part vanishes, and dilaton  obtained as a solution of the \eqn{} \eqref{dilaton_eqn} acquires a rather simple form
\begin{equation*}
\Phi(u)= c_2+c_1\, u+4 \log (\sinh({u})),
\end{equation*}
{where $c_1, c_2$ are arbitrary constants.}

For $|\wt x_4|<1$, the transformation
\begin{align*}\wt x_1 &= v-\frac{1}{2} ({z_3}^2+{z_4}^2) \tanh ({u}), & \wt x_2 &= z_3,\label{xtobrink27tanh} \\
\wt x_4& =\tanh ({u}), & \wt x_3 &= z_4,
\end{align*}
brings the tensor $\wwt F$ into the form
\begin{equation}\label{dual27Bb}
\wt F=   \left(\begin{array}{cccc}
 -2\,\frac{{z_3}^2+{z_4}^2}{ {\cosh}^2({u})} & 1-\tanh (u) & -\frac{{z_3}}{ {\cosh}^2({u})}  &- \frac{{z_4}}{ {\cosh}^2({u})}  \\
 1+\tanh (u) & 0 & 0 & 0 \\
 \frac{{z_3}}{ {\cosh}^2({u})} & 0 & 1 & 0 \\
 \frac{{z_4}}{ {\cosh}^2({u})} & 0 & 0 & 1
\end{array}
\right),
\end{equation}giving the metric\begin{equation}\label{dualmtkBrink27b} ds^{2}= 2dudv-2\,\frac{{z_3}^2+{z_4}^2}{
{\cosh}^2({u})}du^{2}+d{z_3}^2+d{z_4}^2.\end{equation} Torsion again vanishes and dilaton
has the form
   \begin{equation}\Phi(u)= c_2+c_1\, u+4 \log (\cosh ({u})).\end{equation}

We can see that the non-Abelian T-duality w.r.t. the subalgebra $S_{27}$ produces two
types of sigma models in the pp-wave backgrounds, one of them singular and one regular. As
we shall see, this result is obtained also from {dualization w.r.t. }several other
subalgebras of the Poincaré algebra.

\subsubsection{Solution of the classical \eqn s of the \sm}
Our next goal is to write down the general solution of the classical field \eqn s in the backgrounds \eqref{dual27B} and \eqref{dual27Bb}. As their torsions vanish, the antisymmetric parts do  not contribute to the classical field \eqn s.

The Lagrangian for the metric \eqref{dualmtkBrink27} can be written in the form (cf.
\eqref{sma})
\begin{align*}
    \label{dual L27}
L=&\left[ \frac{{Z_3}^2+{Z_4}^2}{ {\sinh}^2({U})} (\partial_{\sigma}U)^2
+\partial_{\sigma}U\partial_{\sigma}V+ \half (\partial_{\sigma}Z_3)^2+\half
(\partial_{\sigma}Z_4)^2\right]
\\&-\left[ \frac{{Z_3}^2+{Z_4}^2}{ {\sinh}^2({U})} (\partial_{\tau}U)^2
+\partial_{\tau}U\partial_{\tau}V+ \half (\partial_{\tau}Z_3)^2+\half
(\partial_{\tau}Z_4)^2\right]. \nonumber
\end{align*}
The field equations then read
\begin{equation}\label{pohrce1}
   \partial^2_{\tau}U- \partial^2_{\sigma}U=0,
\end{equation}
\begin{equation}\label{pohrce3}
    \partial^2_{\sigma}Z_3-\partial^2_{\tau}Z_3=2 \left((\partial_{\sigma}U)^2-(\partial_{\tau}U)^2\right)
   \frac{Z_3}{\sinh^2(U)},
\end{equation}\begin{equation}\label{pohrce4}
    \partial^2_{\sigma}Z_4-\partial^2_{\tau}Z_4=2 \left((\partial_{\sigma}U)^2-(\partial_{\tau}U)^2\right)
   \frac{Z_4}{\sinh^2(U)},
\end{equation}
\begin{eqnarray}
\nonumber \partial^2_{\tau}V-\partial^2_{\sigma}V&=&4\,\text{csch}^2(U)\left[Z_3\,\left(\partial_{\sigma}U\partial_{\sigma}Z_3-\partial_{\tau}U\partial_{\tau}Z_3\right)
\right.\\ && \label{pohrce2} \left. +
Z_4\left(\partial_{\sigma}U \partial_{\sigma}Z_4-\partial_{\tau}U\partial_{\tau}Z_4\right)\right]
    \\ \nonumber
&&    -2\,\text{csch}^3(U) \left[ (Z_3)^2+(Z_4)^2\right]
   \left[(-\partial^2_{\sigma}U +\partial^2_{\tau}U) \sinh (U)\right.\\ && \nonumber \left. +((\partial_{\sigma}U)^2 -(\partial_{\tau}U)^2) \cosh (U)\right].
\end{eqnarray}
To solve these field equations, we can follow steps 1--3 from the section \ref{secPLT}.

Step 1 starts with solution of the field \eqn s in the flat background. In the coordinates
$(t,x,y,z)$ they are of the form \eqref{wavesolns} solved by
$$W^I(\tau,\sigma)=R^I(\tau-\sigma)+L^I(\tau+\sigma),\qquad I=t,x,y,z,$$ with $ R^I,L^I$ arbitrary functions. Subsequent
transformation of this solution to the coordinates $x^\mu$ by using formulas \eqref{txyz27} produces the
functions
\begin{align*}
  X^1(\tau,\sigma) &= -\log(W^t+W^z),&  X^2(\tau,\sigma) &=-\frac{W^x}{W^t+W^z}, \\
 X^4(\tau,\sigma) &= \frac{(W^t)^2-(W^x)^2-(W^y)^2-(W^z)^2}{2(W^t+W^z)}, & X^3(\tau,\sigma) &= \frac{W^y}{W^t+W^z}
\end{align*}
that solve the sigma model field \eqn s  in the flat background in
the coordinates $x^\mu$, i.e. in metric \eqref{mtk27}.

%To get soulution of the classical sigma field \eqn s, first
Next, we have to perform step 2, consisting in the  solution
of the PDEs \eqref{btp2}, \eqref{btm2}, with {Killing fields
\eqref{killings27} on the right-hand sides}. The \eqn s \eqref{btp2}
in this case read
\begin{eqnarray*}
\nonumber  \partial_\tau \tilde h_1 &=& W^t \partial_{\sigma}W^z -W^z \partial_{\sigma}W^t,\\
\label{btp27} \partial_\tau \tilde h_2 &=&-W^x \left(\partial_{\sigma}W^t+\partial_{\sigma}W^z \right)+(W^t+W^z) \partial_{\sigma}W^x,  \\
\nonumber \partial_\tau \tilde h_3 &=&W^y \left(\partial_{\sigma}W^t+\partial_{\sigma}W^z\right)-(W^t+W^z) \partial_{\sigma}W^y,\\
\nonumber \partial_\tau \tilde h_4 &=&
 \partial_{\sigma}W^t +\partial_{\sigma}W^z,
\end{eqnarray*}
while \eqn s \eqref{btm2} are obtained by making the exchange $\tau\leftrightarrow\sigma$.
Compatibility of these two sets of PDEs is guaranteed by the wave \eqn s for $W^I$. Their
solution is
\begin{eqnarray}
\nonumber  \tilde h_1(\tau,\sigma) &=& \gamma_1+\int \left( W^t \partial_{\sigma}W^z -W^z \partial_{\sigma}W^t\right) \, d\tau, \\
\nonumber  \tilde h_2(\tau,\sigma)  &=&\gamma_2-\int \left(W^x \left(\partial_{\sigma}W^t+\partial_{\sigma}W^z \right)-(W^t+W^z) \partial_{\sigma}W^x\right) \, d\tau,  \\
\label{tildehsol27}  \tilde h_3(\tau,\sigma) &=& \gamma_3+\int \left(W^y \left(\partial_{\sigma}W^t+\partial_{\sigma}W^z\right)-(W^t+W^z) \partial_{\sigma}W^y\right) \, d\tau,  \\
\nonumber  \tilde h_4(\tau,\sigma)  &=& \gamma_4+\int \left(\partial_{\sigma}W^t +\partial_{\sigma}W^z\right)
\, d\tau,
\end{eqnarray}where $\gamma_1,\ldots,\gamma_4$ are constants.

To get the solution of field \eqn s \eqref{pohrce1}--\eqref{pohrce2} we have to carry out step 3. One can easily check that the adjoint representation of the algebra
\eqref{comrel27} is faithful, so we can use \eqn {} \eqref{replgh} to express the
coordinates $\wt x_\mu$ in terms of $x^\nu$ and $\wt h_k$. We get
\begin{equation}
    \label{ex272}\wt x_1 = \wt h_1-x^2 \wt h_2-x^3 \wt h_3-x^4 \wt h_4,
\end{equation}
\begin{equation*}
\wt x_2 = e^{x^1}\wt h_2,\qquad
\wt x_3 = e^{x^1} \wt h_3,
   \qquad   \wt x_4 = e^{x^1} \wt h_4.
\end{equation*}
Finally, we have to transform the coordinates $\wt x_\mu$ into the Brinkmann's. Composing the
inverse of \eqref{txyz27}, \eqref{ex272} and the  inverse of \eqref{xtobrink27}, we get the
Brinkmann coordinates $(u,v,z_3,z_4)$ on $\wwt G$ as functions of the spacetime coordinates
$(t,x,y,z)$ on the initial flat manifold and  \coor s $\wt h_j$ on the subgroup $\wwt G $ of
the \dd
\begin{equation}
u = \text{arccoth}\left(\frac{\wt h_4}{t+z}\right),\qquad z_3 = \frac{\wt h_2}{t+z}, \qquad z_4 = \frac{\wt h_3}{t+z}, \label{V27}
\end{equation}
\begin{equation}
\nonumber  v = \frac{\left(2 \wt h_2 x-2 \wt h_3 y+2 \wt h_1 (t+z)+\wt h_4  \left(-t^2+x^2+y^2+z^2\right)\right)}{2 (t+z)}+\frac{\wt h_4 \wt h_2^2+\wt h_3^2 \wt h_4}{2 (t+z)^3}.
\end{equation}

To get {the general} solution of the the classical field \eqn s
\eqref{pohrce1}--\eqref{pohrce2} in the curved background with the metric
\eqref{dualmtkBrink27}, we have to replace the \coor s $(t,x,y,z)$ in \eqref{V27} by the
solutions $W^I=W^I(\tau,\sigma)$ of the wave \eqn s \eqref{wavesolns} and $\wt h_\mu$ by the
solutions
\eqref{tildehsol27} %$\wt h_\mu(\tau,\sigma)$
of the PDEs \eqref{btp2}, \eqref{btm2}. We obtain

\begin{equation}\label{solnpohrce27} U(\tau,\sigma)=\text{arccoth}\left(\frac{\wt h_4(\tau,\sigma)}{W^t(\tau,\sigma)+W^z(\tau,\sigma)}\right),
\end{equation}
\begin{equation*}
 Z_3(\tau,\sigma)=\frac{\wt h_2(\tau,\sigma)}{W^t(\tau,\sigma)+W^z(\tau,\sigma)}, \qquad Z_4(\tau,\sigma)=\frac{\wt h_3(\tau,\sigma)}{W^t(\tau,\sigma)+W^z(\tau,\sigma)}.
\end{equation*}
The expression for the function $V(\tau,\sigma)$ is rather extensive, but can be easily read
out of \eqref{V27}.

{ String-type solutions in the light-cone gauge (see e.g. \cite{papa},\cite{hlapet:ijmp}),
i.e.
\begin{equation}\label{lightcone}  U(\tau,\sigma)=\kappa \tau,\
Z_3(\tau,\sigma)=\sum_{n=-\infty}^{\infty}Z_3^n(\tau)e^{2in\sigma}, \
Z_4(\tau,\sigma)=\sum_{n=-\infty}^{\infty}Z_4^n(\tau)e^{2in\sigma},\end{equation}are
obtained if
$$ W^t(\tau,\sigma)+W^z(\tau,\sigma)=e^{\kappa  \sigma} \sinh (\kappa  \tau),$$
$$W^x(\tau ,\sigma )= \sinh(\kappa \tau)\sum_{n=-\infty}^{\infty}e^{2i n \sigma } (2 i\, n+\kappa)\int Z_3^n(\tau)\text{csch}(\kappa \tau)  d\tau, $$
$$W^y(\tau ,\sigma )= \sinh(\kappa \tau)\sum_{n=-\infty}^{\infty}e^{2i n \sigma } (2 i\, n+\kappa)\int Z_4^n(\tau)\text{csch}(\kappa \tau)  d\tau, $$
where $Z_3^n(\tau)$ and $Z_4^n(\tau)$ solve the differential \eqn }
$${Z}''(\tau )+ \left(4\,n^2-2\,\kappa ^2\text{csch}^2(\kappa \tau)\right) Z(\tau )=0.$$

The solution of the classical field \eqn s in the curved background with the metric
\eqref{dualmtkBrink27b} is obtained from the solution \eqref{solnpohrce27} when
$\text{arccoth}$ is replaced by $\text{arctanh}$.

\subsection{Example 2 -- subalgebra $S_{17}$}\label{example2}

The second example will deal with the subalgebra
$$S_{17} = Span[\mathcal{K}_1= L_3+ \epsilon\,(P_0+ P_3),
\mathcal{K}_2= P_1, \mathcal{K}_3= P_2, \mathcal{K}_4=P_0- P_3] ,\
\epsilon=\pm 1\\$$ which produces a dual model with torsion and whose
representation is not faithful. The commutation relations of this
subalgebra
 are
$$[\mathcal{K}_1,\mathcal{K}_2]=\mathcal{K}_3,\qquad [\mathcal{K}_1,\mathcal{K}_3]=-\mathcal{K}_2.$$
Transformation of \coor s in the flat background
\begin{equation}
\label{txyz17} t =x^1 \epsilon +x^4, \quad  x =x^2, \quad  y =x^3, \quad  z =x^1 \epsilon  -x^4,
\end{equation}
yields components of the flat metric in the group \coor s as
 \begin{equation*}\label{mtz17}
    F_{\mu\nu}(x)=\left(
\begin{array}{cccc}
 0 & 0 & 0 & -2 \epsilon  \\
 0 & 1 & 0 & 0 \\
 0 & 0 & 1 & 0 \\
 -2 \epsilon  & 0 & 0 & 0
\end{array}
\right).
\end{equation*}
The dual background in this case is \begin{equation*}\label{dualmtz17}
    \wwt F_{\mu\nu}(\wt x)=\left(
\begin{array}{cccc}
 0 & 0 & 0 & -\frac{1}{2 \epsilon } \\
 0 & 1 & 0 & \frac{\tilde{x}_3}{2 \epsilon } \\
 0 & 0 & 1 & -\frac{\tilde{x}_2}{2 \epsilon } \\
 -\frac{1}{2 \epsilon } & -\frac{\tilde{x}_3}{2 \epsilon } & \frac{\tilde{x}_2}{2 \epsilon } & -\frac{\tilde{x}_2^2+\tilde{x}_3^2}{4 \epsilon ^2}
\end{array}
\right),
\end{equation*}
and the transformation to Brinkmann \coor s
\begin{equation}\label{BrinkCoors17}\wt x_1=-v,\qquad
\wt x_2=z_3,\qquad \wt x_3=z_4,\qquad \wt x_4=2\epsilon\,u,
\end{equation}
brings the dual metric into the homogeneous and isotropic form
\begin{equation}\label{dualmtkBrink17}
ds^{2}= 2dudv-({z_3}^2+{z_4}^2)\, du^{2}+d{z_3}^2+d{z_4}^2.
\end{equation}
The torsion in Brinkmann \coor s is constant
\begin{equation}\label{torsion17}
H=-2\,du\wedge dz_3\wedge dz_4,
\end{equation}
and the dilaton is
$$ \Phi(u)=c_1+c_2\,u.$$

To find the general solution of the field \eqn s of the dual \sm {} with
torsion, we have to express the coordinates $\wt x_\mu$ in terms of
$x^\nu$ and $\wt h_k$. As the adjoint representation of $S_{17}$ is
not faithful, we have to use the formula \eqref{BHC} to solve the
\eqn\ \eqref{lghtgth} for \coor s of $\wt g$. We get
\begin{align*}
    \wt x_1 &= \wt h_1+x^2 \wt h_3-x^3 \wt h_2,
   & \wt x_2 &=\wt h_2 \cos x^1-\wt h_3\sin x^1,\\
   \wt x_4 &= \wt h_4,
   & \wt x_3 &= \wt h_2\sin x^1+\wt h_3\cos x^1.
\end{align*}

Like in the previous section, combining this with
\eqref{BrinkCoors17} and \eqref{txyz17}, we find the general solution of the
field \eqn s of the \sm\ with metric \eqref{dualmtkBrink17} and
torsion \eqref{torsion17} as
\begin{eqnarray*}\nonumber U(\tau,\sigma)&=&\frac{\wt h_4(\tau,\sigma)}{2\epsilon},\\
V(\tau,\sigma)&=&-\wt h_1(\tau,\sigma)- \wt
h_3(\tau,\sigma)W^x(\tau,\sigma)+ \wt
h_2(\tau,\sigma)W^y(\tau,\sigma),\\ \nonumber
Z_3(\tau,\sigma)&=&\cos(\Omega(\tau,\sigma))\,{\wt h_2(\tau,\sigma)}-\sin(\Omega(\tau,\sigma))\,{\wt h_3(\tau,\sigma)}, \label{solnpohrce17}\\
Z_4(\tau,\sigma)&=&\cos(\Omega(\tau,\sigma)){\wt
h_3(\tau,\sigma)}+\sin(\Omega(\tau,\sigma)){\wt
h_2(\tau,\sigma)},\nonumber
\end{eqnarray*}
where  the $W^I(\tau,\sigma)$ are solutions of the wave \eqn s
\eqref{wavesolns},
$$ \Omega(\tau,\sigma)=\frac{ W^t+ W^z}{2\epsilon },$$ and $\wt
h_\mu$ are solutions of the PDEs \eqref{btp2}, \eqref{btm2},
\begin{equation}
\nonumber \wt h_1 = \gamma_1+\int\Big[
\epsilon\left(\partial_{\tau}W^t -\partial_{\tau}W^z\right)+ W^x
\partial_{\tau}W^y - W^y \partial_{\tau}W^x \Big] \, d\sigma,
\end{equation}
\begin{equation}
\nonumber \wt h_2 = \gamma_2-\int \partial_{\tau}W^x  \, d\sigma,
\qquad  \wt h_3 = \gamma_3-\int \partial_{\tau}W^y  \, d\sigma,
\end{equation}
\begin{equation}
\nonumber \wt h_4 = \gamma_4 +\int
\left(\partial_{\tau}W^t+\partial_{\tau}W^z\right) \, d\sigma.
\end{equation}

{String-type solutions in the light-cone gauge \eqref{lightcone} are obtained if {we
choose}
$$ W^t(\tau,\sigma)+W^z(\tau,\sigma)=2\,\epsilon\,\kappa\,\sigma,$$
\begin{align*}
W^x(\tau ,\sigma )= \sum_{n=-\infty}^{\infty}e^{2i n \sigma } &\int  Z_3^n(\tau) (\kappa
   \sin (\kappa  \sigma )-2i\, n \cos (\kappa  \sigma ))\\
   &-Z_4^n(\tau) (\kappa  \cos (\kappa
   \sigma )+2i\, n \sin (\kappa  \sigma )) \, d\tau,
\end{align*}
\begin{align*}
W^y(\tau ,\sigma )= \sum_{n=-\infty}^{\infty}e^{2i n \sigma } &\int Z_3^n(\tau) (\kappa  \cos (\kappa  \sigma )+2i\, n \sin (\kappa  \sigma
   ))\\ &+Z_4^n(\tau) (\kappa  \sin (\kappa  \sigma )-2i\, n \cos (\kappa  \sigma )) \,
  d\tau.
\end{align*}
where $Z_3^n(\tau)$ and $Z_4^n(\tau)$ solve the system of
differential \eqn s}
$${Z_3^n}''(\tau )+ \left(4\,n^2+\kappa ^2\right) Z_3^n(\tau )-4 i\, n \,\kappa  Z_4^n(\tau )=0,$$
$${Z_4^n}''(\tau ) +\left(4\,n^2+\kappa ^2\right) Z_4^n(\tau )+4 i\, n \,\kappa  Z_3^n(\tau )=0.$$
\subsection{Example 3 -- subalgebra $S_{19}$}\label{example3}

{The third example will deal with the subalgebra
$$  S_{19} = Span[ \mathcal{K}_1=L_3 + \alpha\,P_3, \mathcal{K}_2=P_1, \mathcal{K}_3=P_2,
\mathcal{K}_4=P_0] , \quad\alpha\neq 0$$ which produces diagonalizable dual metric with
nonvanishing scalar curvature and torsion.

The commutation relations of this subalgebra
$$[\mathcal{K}_1,\mathcal{K}_2]=\mathcal{K}_3,\qquad [\mathcal{K}_1,\mathcal{K}_3]=-\mathcal{K}_2$$
are equal to those in the previous example, {but the subalgebras of Killing vectors cannot
be transformed into one another by an element of {the group of proper ortochronous Poincaré
transformations (see \cite{PWZ}),} and the representations of the commutation relations in
Killing vector fields {on $M$} are different.} This leads to a different transformation of
\coor s in the flat background, namely,
\begin{equation}\label{txyz19} x^1=\frac{z}{\alpha},\quad  x^2=x,\quad x^3=y,\quad x^4=t.
\end{equation}
The components of the flat metric in the group
\coor s then read
$$ F_{\mu\nu}=\left(
\begin{array}{cccc}
\alpha ^2 & 0 & 0 & 0 \\
0 & 1 & 0 & 0 \\
0 & 0 & 1 & 0 \\
0 & 0 & 0 & -1
\end{array}\right).$$
The dual background in this case is \begin{equation*}
   \wwt F_{\mu\nu}=\left(
\begin{array}{cccc}
 \frac{1}{\alpha ^2+\wt x_2^2+\wt x_3^2} & \frac{\wt x_3}{\alpha
   ^2+\wt x_2^2+\wt x_3^2} & -\frac{\wt x_2}{\alpha ^2+\wt x_2^2+\wt x_3^2}
   & 0 \\
 -\frac{\wt x_3}{\alpha ^2+\wt x_2^2+\wt x_3^2} & \frac{\alpha
   ^2+\wt x_2^2}{\alpha ^2+\wt x_2^2+\wt x_3^2} & \frac{\wt x_2
   \wt x_3}{\alpha ^2+\wt x_2^2+\wt x_3^2} & 0 \\
 \frac{\wt x_2}{\alpha ^2+\wt x_2^2+\wt x_3^2} & \frac{\wt x_2
   \wt x_3}{\alpha ^2+\wt x_2^2+\wt x_3^2} & \frac{\alpha
   ^2+\wt x_3^2}{\alpha ^2+\wt x_2^2+\wt x_3^2} & 0 \\
 0 & 0 & 0 & -1
\end{array}
\right),
\end{equation*}and its symmetric part gives a metric with nonvanishing scalar curvature
\begin{equation*}
    \wwt R=-\frac{4 \left(\wt x_2^2+\wt x_3^2\right)-10 \alpha
   ^2}{\left(\alpha ^2+\wt x_2^2+\wt x_3^2\right)^2}.
\end{equation*}
This means that it cannot be transformed to the pp-wave form. On the other hand, the
metric of this background can be diagonalized to the form
\begin{equation}\label{dualmtkBrink19}
ds^{2}= -d{y_1}^2+d{y_2}^2+\frac{y_2^2 \alpha ^2}{y_2^2+\alpha
^2}\,d{y_3}^2+\frac{1}{y_2^2+\alpha ^2}\,d{y_4}^2,
\end{equation} via
\begin{equation}\label{BrinkCoors19}\wt
x_1=y_4,\quad \wt x_2=y_2\cos y_3,\quad \wt x_3=y_2\sin y_3, \quad \wt x_4=y_1.
\end{equation}
The torsion then acquires the form
\begin{equation}\label{torsion19}
H=\frac{2 y_2 \alpha ^2}{\left(y_2^2+\alpha
   ^2\right)^2}\,dy_2\wedge dy_3\wedge dy_4,
   \end{equation}
and the dilaton satisfying \eqref{bt1}--\eqref{bt3} is
$$ \Phi=\log(y_2^2+\alpha^2) + const.$$

To find the general solution of the field \eqn s of this dual \sm {}, we have
to express the coordinates $\wt x_\mu$ in terms of $x^\nu$ and $\wt
h_k$. As the adjoint representation of $S_{19}$ is not faithful, we
have to use the formula \eqref{BHC} to solve the \eqn\
\eqref{lghtgth} for the \coor s of $\wt g$. We get
\begin{align*}
    \wt x_1 &= \wt h_1+x^2 \wt h_3-x^3 \wt h_2,
   & \wt x_2 &=\wt h_2 \cos x^1-\wt h_3\sin x^1,\\
   \wt x_4 &= \wt h_4,
   & \wt x_3 &= \wt h_2\sin x^1+\wt h_3\cos x^1.
\end{align*}

Like in the previous section, combining this with
\eqref{BrinkCoors19} and \eqref{txyz19}, we find general solution of the
field \eqn s of the \sm{}  with metric \eqref{dualmtkBrink19} and
torsion \eqref{torsion19} as
\begin{eqnarray*}\nonumber Y_1(\tau,\sigma)&=&\wt h_4(\tau,\sigma),\\
    Y_2(\tau,\sigma)&=&\sqrt{\wt h_2(\tau,\sigma)^2+ \wt
h_3(\tau,\sigma)^2},\\ \nonumber
    Y_3(\tau,\sigma)&=&\arctan\left(\frac{\cos(\Omega(\tau,\sigma)){\wt
h_3(\tau,\sigma)}+\sin(\Omega(\tau,\sigma)){\wt
h_2(\tau,\sigma)}}{\cos(\Omega(\tau,\sigma)){\wt
h_2(\tau,\sigma)}-\sin(\Omega(\tau,\sigma)){\wt
h_3(\tau,\sigma)}}\right), \label{solnpohrce19}\\
    Y_4(\tau,\sigma)&=&\wt h_1(\tau,\sigma)+\wt
h_3(\tau,\sigma)W^x(\tau,\sigma)-\wt
h_2(\tau,\sigma)W^y(\tau,\sigma),\nonumber
\end{eqnarray*}
where  $W^I(\tau,\sigma)$ are solutions of the wave \eqn s
\eqref{wavesolns}, $\Omega(\tau,\sigma)=\frac{W^z(\tau,\sigma)}{\alpha}$, and the $\wt h_\mu$ are solutions of the PDEs
\eqref{btp2},\eqref{btm2},
\begin{equation}
\nonumber \wt h_1 = \gamma_1-\int\Big[ \alpha\,\partial_{\tau}W^z
+W^y\partial_{\tau}W^x-W^x\partial_{\tau}W^y\Big] \, d\sigma,
\end{equation}
\begin{equation}
\nonumber \wt h_2 = \gamma_2-\int \partial_{\tau}W^x  \, d\sigma,
\qquad  \wt h_3 = \gamma_3-\int \partial_{\tau}W^y  \, d\sigma,
\end{equation}
\begin{equation}
\nonumber \wt h_4 = \gamma_4 +\int
\partial_{\tau}W^t \, d\sigma.
\end{equation}
} As this background is not of the pp-wave form, the light-cone gauge cannot be implemented
\cite{HorSteif}. Nevertheless, the field equations are solvable.
\section{Results for other
subalgebras}\label{results} {The classification of subalgebras of
the Poincaré algebra in \cite{PWZ} was carried out  up to the group of
inner automorphisms of the connected component of the Poincaré group
(proper orthochronous Poincaré transformations). There are 35
inequivalent four-dimensional subalgebras of the Poincaré algebra
generated by Killing vectors \eqref{Poincare killings}. }

Only the subgroups corresponding to the subalgebras $S_1$,$S_{2}$,$S_{6}$,$S_{7}$, $S_{8}$,
$S_{11}$, $S_{17}$, $S_{18}$, $S_{19}$, $S_{23}$, $S_{25}-S_{29}$, $S_{31}$, $S_{33}$,
listed in the appendix, %\ref{Poincare subalgebras}
act transitively and freely on the flat spacetime and can be used
for the atomic non-Abelian T-duality. Non-Abelian duals generated
by the algebras $S_1,S_{2}, S_{6}$ {give backgrounds with flat
metric and vanishing torsion. We will not discuss them further.
Dual backgrounds obtained from duality w.r.t. the }subalgebras
$S_{11}$, $S_{18}$, $S_{19}$ have nontrivial scalar curvature. The
others are pp-waves, most of them with nonzero torsion, as we shall
see from the following {list of results. We do not repeat
results for subalgebras $S_{27}$, $S_{17}$, $S_{19}$ described in
section \ref{examples}. }
\subsection{The pp-waves}
\subsubsection{Subalgebras $S_7,S_8$} The
non-isomorphic subalgebras
\begin{eqnarray*}
  S_{7} &=& Span[\mathcal{K}_1=2 M_3 + \alpha\,P_1,\ \mathcal{K}_2= L_2 +M_1,\ \mathcal{K}_3=P_0 - P_3,\ \mathcal{K}_4= P_2], \\
  S_{8} &=& Span[\mathcal{K}_1=M_3 ,\ \mathcal{K}_2= L_2 +M_1,\ \mathcal{K}_3=P_0 - P_3,\ \mathcal{K}_4= P_2],
\end{eqnarray*}
differ only in the value of the parameter $\alpha$ that is positive for $S_{7}$, while
$\alpha=0$ for $S_{8}$ \cite{PWZ}.  The commutation relations of $S_7$ are
$$[\mathcal{K}_1,\mathcal{K}_2]=-2\mathcal{K}_2-\alpha \mathcal{K}_3,\qquad [\mathcal{K}_1,\mathcal{K}_3]=-2\mathcal{K}_3.\ $$ The transformation of \coor s in the flat
background
\begin{align*}
\nonumber  t  &= x^1 x^2 (-\alpha )+\frac{1}{2} e^{-2 x^1} \left((x^2)^2+1\right)+x^3, &
\label{txyz7}  x &=x^1 \alpha -e^{-2 x^1} x^2,\\
\nonumber   z &=x^1 x^2 \alpha -\frac{1}{2} e^{-2 x^1}
   \left((x^2)^2-1\right)-x^3, &  y &=x^4,
\end{align*}
gives components of the flat metric in the group \coor s\begin{equation*}\label{mtz7}
    F_{\mu\nu}(x)=\left(
\begin{array}{cccc}
 \alpha ^2 & -e^{-2 x^1} \alpha  (2 x^1+1) & 2 e^{-2 x^1}
   & 0 \\
 -e^{-2 x^1} \alpha  (2 x^1+1) & e^{-4 x^1} & 0 & 0 \\
 2 e^{-2 x^1} & 0 & 0 & 0 \\
 0 & 0 & 0 & 1
\end{array}
\right).
\end{equation*}
The dual background in this case is
\begin{equation*}\label{dualmtz7}
    \wwt F_{\mu\nu}(\wt x)=\left(
\begin{array}{cccc}
 0 & 0 & \frac{1}{2-2 \wt x_3} & 0 \\
 0 & 1 & \frac{\wt x_3 \alpha +\alpha +2 \wt x_2}{2-2 \wt x_3} & 0 \\
 \frac{1}{2 \wt x_3+2} & \frac{-\wt x_3 \alpha +\alpha -2 \wt x_2}{2 \wt x_3+2} & \frac{(2 \wt x_2+\alpha  \wt x_3)^2}{4 \left(\wt x_3^2-1\right)} & 0 \\
 0 & 0 & 0 & 1
\end{array}
\right),
\end{equation*}
and the torsion vanishes.

The transformation to Brinkmann \coor s {valid} for $|\wt x_3|<1$,
\begin{align*}
&\wt x_1 =-2v -\frac{1}{4} \left( \alpha ^2 u+4 z_3 \alpha +2 z_3 \alpha  \log
   \left(1-\tanh ^2(u)\right)\right)\Big]\\ &+\frac{1}{16} \Big[\tanh (u) \left[\alpha ^2 \log ^2\left(1-\tanh
   ^2(u)\right)+\right. \left. 4 \alpha ^2 \log \left(1-\tanh ^2(u)\right)+16 z_3^2+4
   \alpha ^2\right],
\end{align*}
\begin{equation}\label{BrinkCoors8}
\wt x_2 =  {z_3}-\frac{1}{4} \alpha \tanh (u) \log \left(1-\tanh
^2(u)\right),
\end{equation}
\begin{equation*}
\wt x_3 = -\tanh (u), \qquad  \wt x_4 = z_4.
\end{equation*}
brings the dual metric and dilaton to forms
\begin{equation}\label{dualmtkBrink7}
ds^{2}= 2dudv-2\,\frac{{z_3}^2}{ {\cosh}^2({u})}du^{2}+d{z_3}^2+d{z_4}^2,\end{equation}
   $$ \Phi(u)=c_1+c_2\, u+2 \log (\cosh
   (u)).$$

The transformation for $|\wt x_3|>1$ obtained by replacing $\tanh\rightarrow\coth$ gives dual
metric and dilaton in Brinkmann \coor s
\begin{equation}\label{dualmtkBrink7b}
ds^{2}= 2dudv+2\,\frac{{z_3}^2}{ {\sinh}^2({u})}du^{2}+d{z_3}^2+d{z_4}^2,\end{equation}
$$ \Phi(u)=c_1+c_2\, u+2 \log (\sinh(u)).$$

These results are {\em independent of $\alpha$} and valid for both $S_7$ and $S_8$, hence we can restrict consideration to the simpler case of $S_8$. Even though the adjoint representation of $S_8$ is not faithful, we can solve the \eqn{}
\eqref{lghtgth} for \coor s of $\wt g$
\begin{equation*}
    \wt x_1 = \wt h_1-x^2 \wt h_2-x^3 \wt h_3,
\end{equation*}
\begin{equation}
\label{ex8} \wt x_2 = e^{x^1}\wt h_2,
  \qquad \wt x_3 = e^{x^1} \wt h_3,
   \qquad   \wt x_4 = \wt h_4.
\end{equation}
Like in the previous section, {transformations \eqref{BrinkCoors8} and \eqref{ex8}
enable} us to find the general solution of field \eqn s of the \sm s with metrics
\eqref{dualmtkBrink7} and \eqref{dualmtkBrink7b}.
\subsubsection{Subalgebra $S_{23}$}
\begin{align*}
S_{23} = Span[&\mathcal{K}_1= L_2+M_1-\half\,(P_0+ P_3), \mathcal{K}_2= L_1-M_2+\alpha P_1, \\
&\mathcal{K}_3= P_0- P_3,
\mathcal{K}_4=P_2],\quad \alpha>0.
\end{align*}
The commutation relations are
$$[\mathcal{K}_1,\mathcal{K}_2]=\alpha\,\mathcal{K}_3-\mathcal{K}_4,\qquad [\mathcal{K}_2,\mathcal{K}_4]=-\mathcal{K}_3,\qquad \alpha > 0.\ $$ The transformation of \coor s in the flat
background is
\begin{align*}
t  &=\frac{1}{6} \left(-(x^1)^3-3 \left((x^2)^2+1\right)
   x^1+6 x^3\right), \label{txyz23}  & x &=x^2 \alpha +\frac{(x^1)^2}{2} ,\\
\nonumber   z &=\frac{1}{6} \left((x^1)^3+3 \left((x^2)^2-1\right)
   x^1-6 x^3\right), &  y &=x^4-x^1
   x^2.
\end{align*}
The flat metric in the group \coor s reads
\begin{equation*}\label{mtz23}
    F_{\mu\nu}(x)=\left(
\begin{array}{cccc}
 0 & \alpha  x^1 & 1 & -x^2 \\
 \alpha  x^1 & \alpha ^2+(x^1)^2 & 0 & -x^1 \\
 1 & 0 & 0 & 0 \\
 -x^2 & -x^1 & 0 & 1
\end{array}
\right),
\end{equation*}
and the dual background is given by
\begin{equation*}\label{dualmtz23}
    \wwt F_{\mu\nu}(\wt x)=\left(
\begin{array}{cccc}
 0 & 0 & 1 & 0 \\
 0 & \frac{1}{\alpha ^2+ \wt x_3^2} & \frac{ \wt x_4-\alpha
    \wt x_3}{\alpha ^2+ \wt x_3^2} & -\frac{ \wt x_3}{\alpha ^2+ \wt x_3^2}
   \\
 1 & \frac{\alpha   \wt x_3- \wt x_4}{\alpha ^2+ \wt x_3^2} &
   -\frac{( \wt x_4-\alpha   \wt x_3)^2}{\alpha ^2+ \wt x_3^2} &
   \frac{ \wt x_3 ( \wt x_4-\alpha   \wt x_3)}{\alpha ^2+ \wt x_3^2} \\
 0 & \frac{ \wt x_3}{\alpha ^2+ \wt x_3^2} & \frac{ \wt x_3 ( \wt x_4-\alpha
    \wt x_3)}{\alpha ^2+ \wt x_3^2} & \frac{\alpha ^2}{\alpha
   ^2+ \wt x_3^2}
\end{array}
\right).
\end{equation*}
The dual metric in Brinkmann \coor s
\begin{align*}
\wt x_1&=\frac{1}{24 \left(1+u^2\right)^{3/2} \alpha }\left(-12 \left(-4+u^2\right) \left(1+u^2\right)^2 \alpha ^2 z_3-12 u \sqrt{1+u^2} \left(2+u^2\right) z_3^2\right.\\ &+\left. \sqrt{1+u^2} \left(\left(1+u^2\right) \left(24 v+u \left(-48+28 u^2-3 u^4\right) \alpha ^4\right)-12 u z_4^2\right)\right)
,\\
\wt x_2&=\sqrt{1+u^2} \alpha  z_4, \qquad \wt x_3=u \alpha ,\qquad \wt x_4=\frac{1}{2} u \left(-4+u^2\right) \alpha ^2+\sqrt{1+u^2} z_3
\end{align*}
then has the form
\begin{equation*}\label{dualmtkBrink23}
ds^{2}= 2dudv+\frac{\left(2 u^2-1\right) {z_4}^2-3
   {z_3}^2}{\left(u^2+1\right)^2}du^{2}+d{z_3}^2+d{z_4}^2,
\end{equation*}
while the torsion and the dilaton are
$$ H=\frac{2}{1+u^2}\,du\wedge dz_3\wedge dz_4,\qquad \Phi(u)=c_1+c_2\,u+\log(1+u^2). $$
To find the general solution of field \eqn s of the
dual \sm {}, we have to express the coordinates $\wt x_\mu$ in terms of $x^\nu$ and $\wt
h_k$. We get
\begin{align*}
    \wt x_1 &= \wt h_1+x^2 (\alpha\,\wt h_3-\half x^2\wt h_3-\wt h_4), & \wt x_3 &= \wt h_3,   \\
   \wt x_2 &=  \wt h_2-x^1 (\alpha\,\wt h_3- x^2\wt h_3-\wt h_4), &  \wt x_4 &= x^2\wt h_3+\wt h_4.
\end{align*}

\subsubsection{Subalgebra $S_{25}$}
$$S_{25} = Span[\mathcal{K}_1= L_2+M_1-\epsilon\,P_2, \mathcal{K}_2=P_0+ P_3, \mathcal{K}_3=P_1, \mathcal{K}_4 =P_0- P_3],
\ \epsilon=\pm 1. $$ The commutation relations are
$$[\mathcal{K}_1,\mathcal{K}_2]=2\,\mathcal{K}_3,\qquad [\mathcal{K}_1,\mathcal{K}_3]=\mathcal{K}_4.$$ The transformation of \coor s in the flat
background
\begin{equation*}
\label{txyz25} t=x^2+x^4,\quad  x =x^3 ,
\quad   y =-\epsilon\, x^1, \quad   z =x^2-x^4,
\end{equation*}
{yields the flat metric in the group \coor s}
\begin{equation*}\label{mtz25}
    F_{\mu\nu}(x)=\left(
\begin{array}{cccc}
 \epsilon^2 & 0 & 0 & 0 \\
 0 & 0 & 0 & -2 \\
 0 & 0 & 1 & 0 \\
 0 & -2 & 0 & 0
\end{array}
\right).
\end{equation*}
The dual background {is given by}
\begin{equation*}\label{dualmtz25}
    \wwt F_{\mu\nu}(\wt x)=\left(
\begin{array}{cccc}
 \frac{1}{ \wt x_4^2+\epsilon^2} & 0 & \frac{ \wt x_4}{ \wt x_4^2+\epsilon^2} &
   -\frac{ \wt x_3}{ \wt x_4^2+\epsilon^2} \\
 0 & 0 & 0 & -\frac{1}{2} \\
 -\frac{ \wt x_4}{ \wt x_4^2+\epsilon^2} & 0 & \frac{\epsilon^2}{ \wt x_4^2+\epsilon^2} &
   \frac{ \wt x_3  \wt x_4}{ \wt x_4^2+\epsilon^2} \\
 \frac{ \wt x_3}{ \wt x_4^2+\epsilon^2} & -\frac{1}{2} & \frac{ \wt x_3
    \wt x_4}{ \wt x_4^2+\epsilon^2} & -\frac{ \wt x_3^2}{ \wt x_4^2+\epsilon^2}
\end{array}
\right).
\end{equation*}
The transformation to Brinkmann \coor s
\begin{align*}
  \wt x_1 &= \epsilon\, \sqrt{u^2+1}\, z_4,  & \wt x_3 &= \sqrt{u^2+1}\, z_3,\\
  \wt x_2 &= \frac{1}{\epsilon(u^2+1)}\left[u \left(u^2+2\right)
   z_3^2+u z_4^2\right]-{2\epsilon\, v}, &
  \wt x_4 &=  \epsilon\, u.
\end{align*}
{gives the dual metric}
\begin{equation*}\label{dualmtkBrink25}
ds^{2}= 2dudv+\frac{\left(2 {u}^2-1\right) {z_4}^2-3
   {z_3}^2}{\left({u}^2+1\right)^2}du^{2}+d{z_3}^2+d{z_4}^2.
\end{equation*}
The torsion and dilaton then read
$$ H=-\frac{2}{1+u^2}\,du\wedge dz_3\wedge dz_4, \qquad \Phi(u)=c_1+c_2\,u+\log(1+u^2).$$
To find the general solution of field \eqn s of the dual \sm {}, we have to express the coordinates
$\wt x_\mu$ in terms of $x^\nu$ and $\wt h_k$. We get
\begin{align*}
    \wt x_1 &= \wt h_1+2 x^2 \wt h_3+ x^3\wt h_4, &\wt x_3 &= \wt h_3- x^1\wt h_4, \\
   \wt x_2 &=  \wt h_2-x^1 (2\wt h_3- x^1\wt h_4),&
    \wt x_4 &= \wt h_4.
\end{align*}
\subsubsection{Subalgebras $S_{26},\ S_{27}$}
The subalgebras $S_{26},S_{27}$ differ once again only in the value of the parameter $\alpha$: it {is positive for $S_{26}$,} while $\alpha=0$ for $S_{27}$ \cite{PWZ}.
\begin{eqnarray*}
 S_{26} &=& Span[\mathcal{K}_1= M_3+\alpha P_1, \mathcal{K}_2= L_2+M_1, \mathcal{K}_3=L_1-M_2, \mathcal{K}_4 =P_0- P_3],\\
 S_{27} &=& Span[\mathcal{K}_1= M_3, \mathcal{K}_2= L_2+M_1, \mathcal{K}_3=L_1-M_2, \mathcal{K}_4 =P_0- P_3].
\end{eqnarray*}
Their commutation relations are
$$[\mathcal{K}_1,\mathcal{K}_2]=-\mathcal{K}_2-\alpha \mathcal{K}_4,\qquad [\mathcal{K}_1,\mathcal{K}_3]=-\mathcal{K}_3,\qquad  [\mathcal{K}_1,\mathcal{K}_4]=-\mathcal{K}_4.
$$
The transformation of \coor s in the flat background
\begin{align*}
t  &= -\alpha\,x^1 x^2 +\frac{1}{2} e^{-x^1} \left((x^2)^2+(x^3)^2+1\right)+x^4 ,& x &= \alpha\,x^1 -e^{-x^1} x^2,\label{txyz26} \\
z &=\alpha\,x^1 x^2 -\frac{1}{2} e^{-x^1} \left((x^2)^2+(x^3)^2-1\right)-x^4, & y &= e^{-x^1} x^3
\end{align*}
{gives the flat} metric in the group \coor s
\begin{equation*}\label{mtz26}
    F_{\mu\nu}(x)=\left(
\begin{array}{cccc}
 \alpha ^2 & -e^{-x^1} \alpha  (x^1+1) & 0 & e^{-x^1} \\
 -e^{-x^1} \alpha  (x^1+1) & e^{-2 x^1} & 0 & 0 \\
 0 & 0 & e^{-2 x^1} & 0 \\
 e^{-x^1} & 0 & 0 & 0
\end{array}
\right).
\end{equation*}
{In the dual background
\begin{equation*}\label{dualmtz26}
    \wwt F_{\mu\nu}(\wt x)=\left(
\begin{array}{cccc}
 0 & 0 & 0 & \frac{1}{1- \wt x_4} \\
 0 & 1 & 0 & \frac{ \wt x_4 \alpha +\alpha + \wt x_2}{1- \wt x_4} \\
 0 & 0 & 1 & \frac{ \wt x_3}{1- \wt x_4} \\
 \frac{1}{ \wt x_4+1} & \frac{- \wt x_4 \alpha +\alpha - \wt x_2}{ \wt x_4+1} &
   -\frac{ \wt x_3}{ \wt x_4+1} & \frac{ \wt x_2^2+2 \alpha   \wt x_4  \wt x_2+ \wt x_3^2+\alpha ^2
    \wt x_4^2}{ \wt x_4^2-1}
\end{array}
\right)
\end{equation*}
the torsion vanishes.}

The transformation to Brinkmann \coor s  \begin{eqnarray*}
  \wt x_1 &=& -v+\frac{1}{8} \Big(-4 u \alpha ^2+\tanh (u)
  \left(4 \left(z_3^2+z_4^2+\alpha
^2\right)+\right.\\ && \left.\alpha ^2 \log \left(1-\tanh^2(u)\right) \left(\log
\left(1-\tanh ^2(u)\right)+4\right)\right)- \\ && 4 z_3 \alpha  \left(\log \left(1-\tanh
   ^2(u)\right)+2\right)\Big),\\
  \wt x_2 &=& z_3-\frac{1}{2} \alpha  \tanh (u) \log \left(1-\tanh ^2(u)\right),\\
  \wt x_3 &=& z_4,\\  \wt x_4 &=&  -\tanh(u),
\end{eqnarray*}
for $|\wt x_1|<1$, brings the dual metric and dilaton  to forms independent of $\alpha$
\begin{equation*}\label{dualmtkBrink26}
ds^{2}= 2dudv-2\,\frac{{z_3}^2+{z_4}^2}{ {\cosh}^2({u})}du^{2}+d{z_3}^2+d{z_4}^2,
\end{equation*}
$$ \Phi(u)=c_1+c_2\, u +4 \log (\cosh(u)).$$

A similar transformation (see Sec. \ref{example}) gives the dual metric and dilaton for $|\wt
x_1|>1$ in Brinkmann \coor s
\begin{equation*}\label{dualmtkBrink26b}
ds^{2}= 2dudv+2\,\frac{{z_3}^2+{z_4}^2}{ {\sinh}^2({u})}du^{2}+d{z_3}^2+d{z_4}^2,
\end{equation*}
$$ \Phi(u)=c_1+c_2\, {u}+4 \log (\sinh({u})).$$

The solution of the field \eqn s of the dual \sm s was found in Sec. \ref{example}.
\subsubsection{Subalgebra $S_{28}$}
$$S_{28} = Span[\mathcal{K}_1=L_3 - \beta\,M_3, \mathcal{K}_2=L_2 + M_1, \mathcal{K}_3=L_1 - M_2, \mathcal{K}_4=P_0 - P_3] , \quad  \beta\neq 0.$$
The commutation relations are
$$[\mathcal{K}_1,\mathcal{K}_2]=\beta\,\mathcal{K}_2-\mathcal{K}_3,\qquad [\mathcal{K}_1,\mathcal{K}_3]=\,\mathcal{K}_2+\beta\,\mathcal{K}_3,\qquad
[\mathcal{K}_1,\mathcal{K}_4]=\beta\,\mathcal{K}_4,\qquad \ \beta\neq 0.$$
The transformation of \coor s in the flat background
\begin{align*}
  t  &=\frac{1}{2} \left((x^2)^2+(x^3)^2+1\right) e^{x^1 \beta }+x^4,&  x & =  x^2 \left(-e^{x^1 \beta }\right),\\
z &=-\frac{1}{2} \left((x^2)^2+(x^3)^2-1\right) e^{x^1 \beta }-x^4  ,& y &= x^3 e^{x^1 \beta },
\end{align*}
gives the flat metric in the group \coor s
\begin{equation*}\label{mtz28}
    F_{\mu\nu}(x)=\left(
\begin{array}{cccc}
 0 & 0 & 0 & -e^{\beta\,x^1}\beta \\
 0 & e^{2\beta\, x^1} & 0 & 0 \\
 0 & 0 & e^{2\beta\, x^1} & 0 \\
 -e^{\beta\,x^1}\beta & 0 & 0 & 0
\end{array}
\right).
\end{equation*}
{After the transformation
\begin{eqnarray*}\wt x_1&=&\frac{1}{2} \beta  \left(2 v-\tanh (u)
\left({z_3}^2+{z_4}^2\right)\right),
\\ \wt x_2&=&{z_3} \cos \left(\frac{\log
(\cosh
   (u))}{\beta }\right)+{z_4} \sin \left(\frac{\log (\cosh (u))}{\beta }\right),
\\   \wt x_3&=&{z_4} \cos \left(\frac{\log (\cosh (u))}{\beta}\right)
-{z_3} \sin \left(\frac{\log (\cosh (u))}{\beta }\right),
\\   \wt x_4&=&-\tanh (u)
\end{eqnarray*}   of the dual background
\begin{equation*}\label{dualmtz28}
    \wwt F_{\mu\nu}(\wt x)=\left(
\begin{array}{cccc}
 0 & 0 & 0 & \frac{1}{\beta  ( \wt x_4-1)} \\
 0 & 1 & 0 & \frac{\beta   \wt x_2- \wt x_3}{\beta -\beta   \wt x_4} \\
 0 & 0 & 1 & \frac{ \wt x_2+\beta   \wt x_3}{\beta -\beta   \wt x_4} \\
 -\frac{1}{ \wt x_4 \beta +\beta } & \frac{ \wt x_3-\beta   \wt x_2}{\beta  ( \wt x_4+1)} &
   -\frac{ \wt x_2+\beta   \wt x_3}{ \wt x_4 \beta +\beta } & \frac{\left(\beta ^2+1\right)
   \left( \wt x_2^2+ \wt x_3^2\right)}{\beta ^2 \left( \wt x_4^2-1\right)}
\end{array}
\right)
\end{equation*}
the dual metric and the dilaton for $|\wt x_1|<1$ are expressed in the Brinkmann \coor s as
}
\begin{equation*}\label{dualmtkBrink28}
ds^{2}= 2dudv-\frac{\left(z_3^2+z_4^2\right)
 \left(1+2 \beta ^2 \text{sech}^2(u)\right) }{ \beta ^2} du^{2}+d{z_3}^2+d{z_4}^2,
 \end{equation*}
   $$ \Phi(u)=c_1+c_2\, u +4 \log (\cosh(u)).$$
The dual metric and the dilaton for $|\wt x_1|>1$ in Brinkmann \coor s are
\begin{equation*}\label{dualmtkBrink28b}
ds^{2}= 2dudv-\frac{\left(z_3^2+z_4^2\right)
 \left(1-2 \beta ^2 \text{csch}^2(u)\right) }{ \beta ^2} du^{2}+d{z_3}^2+d{z_4}^2,
 \end{equation*}
$$ \Phi(u)=c_1+c_2\, {u}+4 \log (\sinh({u})).$$
In both cases the torsion is of the form
\begin{equation*}\label{torsion28}
   H= -\frac{2}{ \beta}\,du\wedge dz_3\wedge dz_4.
\end{equation*}

To find the solution of the \eqn s of the dual \sm {}, we also need $\wt x_j, h^k$ expressed
in terms of $x^j$ as
\begin{eqnarray*}
   \nonumber \wt x_1 &=& x^2 \wt h_2 \beta +x^3 \wt h_3 \beta +x^4 \wt h_4 \beta
   +\wt h_1+x^3 \wt h_2-x^2 \wt h_3,
   \\ \label{tx282}  \wt x_2 &=& e^{-\beta\wt x_1} (\wt h_3 \sin (x^1)+\wt h_2 \cos(x^1)),
   \\ \nonumber  \wt x_3 &=& e^{-\beta\wt x_1} (-\wt h_2 \sin (x^1)+\wt h_3 \cos (x^1)),
   \\ \nonumber   \wt x_4 &=& \wt h_4 e^{{-\beta\wt x_1}} \label{tx280} .
\end{eqnarray*}

\subsubsection{Subalgebra $S_{29}$}
$$S_{29} = Span[\mathcal{K}_1=L_3 - \beta\,M_3, \mathcal{K}_2=P_0 - P_3, \mathcal{K}_3=P_1, \mathcal{K}_4=P_2] ,\quad  \beta\neq 0.$$
The commutation relations are
$$[\mathcal{K}_1,\mathcal{K}_2]=\beta\,\mathcal{K}_2,\qquad [\mathcal{K}_1,\mathcal{K}_3]=\mathcal{K}_4,\qquad [\mathcal{K}_1,\mathcal{K}_4]=-\mathcal{K}_3,\qquad \ \beta\neq 0.$$
The transformation of \coor s in the flat background
\begin{align*}
  t= -\frac{1}{2} e^{x^1 \beta
   }+x^2,\quad  x = x^3,\quad   y = x^4, \quad z = -\frac{1}{2} \left(e^{x^1 \beta }\right)-x^2
\end{align*}
gives the flat metric in the group \coor s
\begin{equation*}\label{mtz29}
    F_{\mu\nu}(x)=\left(
\begin{array}{cccc}
 0 & e^{\beta  x^1} \beta  & 0 & 0 \\
 e^{\beta  x^1} \beta  & 0 & 0 & 0 \\
 0 & 0 & 1 & 0 \\
 0 & 0 & 0 & 1
\end{array}
\right).
\end{equation*}
The dual background is
\begin{equation*}\label{dualmtz29}
    \wwt F_{\mu\nu}(\wt x)=\left(
\begin{array}{cccc}
 0 & \frac{1}{ \wt x_2 \beta +\beta } & 0 & 0 \\
 \frac{1}{\beta -\beta   \wt x_2} & \frac{ \wt x_3^2+ \wt x_4^2}{\beta ^2 \left( \wt x_2^2-1\right)} & \frac{ \wt x_4}{\beta -\beta   \wt x_2} &
   \frac{ \wt x_3}{\beta  ( \wt x_2-1)} \\
 0 & -\frac{ \wt x_4}{ \wt x_2 \beta +\beta } & 1 & 0 \\
 0 & \frac{ \wt x_3}{ \wt x_2 \beta +\beta } & 0 & 1
\end{array}
\right).
\end{equation*}
The dual metric, dilaton and torsion in Brinkmann \coor s are the same as in section \ref{example2}
\begin{equation*}\label{dualmtkBrink29b}
ds^{2}= 2dudv-\left({z_3}^2+{z_4}^2\right)
 du^{2}+d{z_3}^2+d{z_4}^2,
\end{equation*}
\begin{equation*}\label{torsion29}
\Phi(u)=c_1+c_2\, {u}, \qquad
   H= -2\,du\wedge dz_3\wedge
   dz_4.
\end{equation*}

To find the solution of the \eqn s of motion of the dual \sm {}, we also need $\wt x_j, h^k$
expressed in terms of $x^j$ as
\begin{align*}
   \wt x_1 &=x^2 \wt h_2 \beta +\wt h_1-x^4 \wt h_3+x^3 \wt h_4, & \wt x_3 &= \wt h_3 \cos (x^1)-\wt h_4 \sin (x^1),\\
%   \label{tx292}
    \wt x_2 &=\wt h_2 e^{x^1(-\beta )}, & \wt x_4 &= \wt h_3\sin (x^1)+\wt h_4\cos(x^1).
\end{align*}

\subsubsection{Subalgebras $S_{31}, S_{33}$}
The subalgebras
\begin{eqnarray*}
 S_{31} &=& Span[\mathcal{K}_1=M_3, \mathcal{K}_2=P_1 + \beta\,P_2, \mathcal{K}_3=P_0 - P_3, \mathcal{K}_4=L_2 + M_1] ,\\
  S_{33} &=& Span[\mathcal{K}_1=M_3 + \alpha\,P_2,  \mathcal{K}_2=P_1 + \beta\,P_2, \mathcal{K}_3=P_0 - P_3, \mathcal{K}_4=L_2 + M_1],
  \end{eqnarray*}
 differ only in the value of the parameter $\alpha$: {it is positive for $S_{33}$, while} $\alpha=0$ for $S_{31}$. In both cases $\beta \neq 0$. The
subalgebras are isomorphic even though they are not equivalent under conjugacy through proper
orthochronous Poincaré transformations.  Their commutation relations are
$$[\mathcal{K}_1,\mathcal{K}_3]=-\mathcal{K}_3,\qquad [\mathcal{K}_1,\mathcal{K}_4]=-\mathcal{K}_4,\qquad [\mathcal{K}_2,\mathcal{K}_4]=-\mathcal{K}_3.$$ The transformation of \coor s in the flat
background
\begin{align*}
t  &=x^3-x^2 x^4-\frac{1}{2} e^{-x^1}
   \left((x^4)^2+1\right), \label{txyz33}&  x &=x^2+e^{-x^1} x^4,\\
\nonumber   z &=-x^3+x^2 x^4+\frac{1}{2} e^{-x^1}
   \left((x^4)^2-1\right), & y &=x^1 \alpha
   +x^2 \beta ,
\end{align*}
gives the flat metric in the group \coor s as
\begin{equation*}\label{mtz33}
    F_{\mu\nu}(x)=\left(
\begin{array}{cccc}
 \alpha ^2 & \alpha  \beta  & -e^{-x^1} & e^{-x^1} x^2 \\
 \alpha  \beta  & \beta ^2+1 & 0 & e^{-x^1} \\
 -e^{-x^1} & 0 & 0 & 0 \\
 e^{-x^1} x^2 & e^{-x^1} & 0 & e^{-2 x^1}
\end{array}
\right).
\end{equation*}
{For the dual background
\begin{equation*}\label{dualmtz33}
    \wwt F_{\mu\nu}(\wt x)=\left(
\begin{array}{cccc}
 0 & 0 & -\frac{1}{ \wt x_3+1} & 0 \\
 0 & \frac{1}{\beta ^2+ \wt x_3^2} & \frac{\alpha  \beta
   +( \wt x_3+1)  \wt x_4}{( \wt x_3+1) \left(\beta ^2+ \wt x_3^2\right)} &
   -\frac{ \wt x_3+1}{\beta ^2+ \wt x_3^2} \\
 \frac{1}{ \wt x_3-1} & \frac{-\alpha  \beta - \wt x_3
    \wt x_4+ \wt x_4}{( \wt x_3-1) \left(\beta ^2+ \wt x_3^2\right)} &
   \frac{\alpha ^2  \wt x_3^2-2 \alpha  \beta   \wt x_4
    \wt x_3+\left(\beta ^2+1\right)
    \wt x_4^2}{\left( \wt x_3^2-1\right) \left(\beta
   ^2+ \wt x_3^2\right)} & \frac{\alpha  \beta
   ( \wt x_3+1)-\left(\beta ^2+1\right)  \wt x_4}{( \wt x_3-1)
   \left(\beta ^2+ \wt x_3^2\right)} \\
 0 & \frac{ \wt x_3-1}{\beta ^2+ \wt x_3^2} & \frac{\alpha  \beta
   ( \wt x_3-1)-\left(\beta ^2+1\right)  \wt x_4}{( \wt x_3+1)
   \left(\beta ^2+ \wt x_3^2\right)} & \frac{\beta ^2+1}{\beta
   ^2+ \wt x_3^2}
\end{array}
\right),
\end{equation*}
we can find a rather complicated coordinate transformation that enables us to eliminate the $\alpha$-dependence of
the background.} The dual metric, torsion and dilaton for $|x_3|<1$ in Brinkmann \coor s are
\begin{eqnarray*}\label{dualmtzBrink33b}
ds^{2}&=& 2dudv+\left[{z_3}^2\frac{\text{sech}^4(u) \left(2 \left(\beta ^2+1\right)
\sinh ^2(u)-\beta ^2\right)}{\left(\tanh ^2(u)+\beta ^2\right)^2} \right. \\
&&\nonumber  \left. -{z_4}^2\frac{\beta ^2 \text{sech}^4({u}) \left( 2 (\beta ^2+1)
\cosh^2({u})+1\right)}{\left(\tanh^2({u})+\beta ^2\right)^2} \right]du^2+ d{z_3}^2+d{z_4}^2,
\end{eqnarray*}
$$H= \frac{2 \beta}{\beta ^2\cosh^2({u})+\sinh^2({u})}\,du\wedge dz_3\wedge
   dz_4,$$
$$ \Phi(u)=c_1+c_2\,u+\log \left(\left(\beta ^2+1\right) \cosh (2 u)+\beta ^2-1\right).$$
The dual metric, torsion and dilaton for $|x_3|>1$ in Brinkmann \coor s are
\begin{eqnarray*}\label{dualmtzBrink33}
ds^{2}&=& 2dudv+\left[{z_4}^2\frac{\beta ^2 \text{csch}^4(u) \left(2(\beta ^2+1)\sinh ^2(u)-1\right)}{\left(\coth ^2(u)+\beta ^2\right)^2} \right. \\
&&\nonumber  \left.  -{z_3}^2\frac{\text{csch}^4(u) \left(2(\beta ^2+1)\cosh ^2(u)+\beta ^2 \right)}{\left(\coth ^2(u)+\beta ^2\right)^2}\right]du^2+
d{z_3}^2+d{z_4}^2,
\end{eqnarray*}
$$
   H= -\frac{2 \beta}{\cosh^2({u})+\beta ^2\sinh^2({u})}\,du\wedge dz_3\wedge
   dz_4,
$$
$$ \Phi(u)=c_1+c_2\,u+\log \left(\left(\beta ^2+1\right) \cosh (2 u)-\beta ^2+1\right).$$
To find the general solution of field \eqn s of dual \sm {}, it is sufficient to express the
coordinates $\wt x_\mu$ in terms of $x^\nu$ and $\wt h_k$ {for $\alpha=0$}. We get
\begin{align*}
    \wt x_1 &= \wt h_1- x^3 \wt h_3- x^4\wt h_4, & \wt x_2 &=  \wt h_2-x^4\wt h_3,\\
   \wt x_3 &= e^{x^1}\wt h_3, & \wt x_4 &= e^{x^1}(x^2\wt h_3+\wt h_4).
\end{align*}
\subsection{Diagonalizable metrics with nontrivial scalar curvature}
\subsubsection{Subalgebra $S_{11}$}
{
$$S_{11} = Span[\mathcal{K}_1= M_3 + \alpha\,P_2, \mathcal{K}_2=  P_0, \mathcal{K}_3= P_3,
\mathcal{K}_4=P_1],\quad \alpha>0\\.$$ The commutation relations are
$$[\mathcal{K}_1,\mathcal{K}_2]=\mathcal{K}_3,\qquad [\mathcal{K}_1,\mathcal{K}_3]=\mathcal{K}_2.$$
The flat metric in the group \coor s $$x^1=\frac{y}{\alpha},\quad  x^2=t,\quad  x^3=z,\quad
x^4=x$$ reads
$$ F_{\mu\nu}=\left(
\begin{array}{cccc}
\alpha ^2 & 0 & 0 & 0 \\
0 & -1 & 0 & 0 \\
0 & 0 & 1 & 0 \\
0 & 0 & 0 & 1
\end{array}\right).$$
The dual background  \begin{equation*}
   \wwt F_{\mu\nu}=\left(
\begin{array}{cccc}
 \frac{1}{\alpha ^2+\wt x_2^2-\wt x_3^2} & -\frac{\wt x_3}{\alpha
   ^2+\wt x_2^2-\wt x_3^2} & \frac{\wt x_2}{\alpha ^2+\wt x_2^2-\wt x_3^2} &
   0 \\
 \frac{\wt x_3}{\alpha ^2+\wt x_2^2-\wt x_3^2} & -\frac{\alpha
   ^2+\wt x_2^2}{\alpha ^2+\wt x_2^2-\wt x_3^2} & \frac{\wt x_2
   \wt x_3}{\alpha ^2+\wt x_2^2-\wt x_3^2} & 0 \\
 -\frac{\wt x_2}{\alpha ^2+\wt x_2^2-\wt x_3^2} & \frac{\wt x_2
   \wt x_3}{\alpha ^2+\wt x_2^2-\wt x_3^2} & \frac{\alpha
   ^2-\wt x_3^2}{\alpha ^2+\wt x_2^2-\wt x_3^2} & 0 \\
 0 & 0 & 0 & 1
\end{array}
\right)
\end{equation*}gives a metric with nonvanishing scalar curvature
\begin{equation*}
    \wwt R=\frac{2 \left(2 \wt x_2^2-2 \wt x_3^2-5 \alpha
   ^2\right)}{\left(\wt x_2^2-\wt x_3^2+\alpha ^2\right)^2},
\end{equation*}so that it cannot be transformed to the pp-wave form. On the other hand, the
metric of this background can be diagonalized to the time-dependent
form
\begin{equation}\label{dualmtkBrink11}
ds^{2}= -d{y_1}^2+d{y_2}^2+\frac{y_1^2 \alpha ^2}{y_1^2+\alpha
^2}\,d{y_3}^2+\frac{1}{y_1^2+\alpha ^2}\,d{y_4}^2,
\end{equation} via
\begin{equation*}\label{BrinkCoors11}\wt
x_1=y_4,\quad \wt x_2=y_1\cosh y_3,\quad \wt x_3=y_1\sinh y_3,\quad \wt x_4=y_2.
\end{equation*}
The torsion then acquires the form
$$ H=-\frac{2 y_1 \alpha ^2}{\left(y_1^2+\alpha
   ^2\right)^2}\,dy_1\wedge dy_3\wedge dy_4,$$
and the dilaton satisfying \eqref{bt1}--\eqref{bt3} is
$$ \Phi=\log(y_1^2+\alpha^2) + const.$$

{To find the solution of the \eqn s of this dual \sm {}, we need the above \tfn{} between
$y_j$ and $\wt x_j$, and also $\wt x_j$ expressed in terms of $x^j,\wt h_k$.}
\begin{align*}
    \wt x_1 &= \wt h_1+x^2 \wt h_3+x^3 \wt h_2,
   & \wt x_2 &=\wt h_2 \cosh x^1-\wt h_3\sinh x^1,\\
   \wt x_4 &= \wt h_4,
   & \wt x_3 &= \wt h_3\cosh x^1-\wt h_2\sinh x^1.
\end{align*}
\subsubsection{Subalgebra $S_{18}$}
$$ S_{18} = Span[\mathcal{K}_1= L_3 + \alpha\,P_0, \mathcal{K}_2=P_1,
\mathcal{K}_3=P_2, \mathcal{K}_4=P_3] ,\quad \alpha>0.$$ The commutation relations are {the
same as for $S_{17}$ and $S_{19}$,}
$$[\mathcal{K}_1,\mathcal{K}_2]=\mathcal{K}_3,\qquad [\mathcal{K}_1,\mathcal{K}_3]=-\mathcal{K}_2.$$
The flat metric in the group \coor s $$x^1=\frac{t}{\alpha} ,\quad  x^2=x,\quad  x^3=y,\quad
x^4=z$$ reads
$$ F_{\mu\nu}=\left(
\begin{array}{cccc}
-\alpha ^2 & 0 & 0 & 0 \\
0 & 1 & 0 & 0 \\
0 & 0 & 1 & 0 \\
0 & 0 & 0 & 1
\end{array}\right).$$
The dual background  \begin{equation*}
   \wwt F_{\mu\nu}=\left(
\begin{array}{cccc}
 \frac{1}{-\alpha ^2+\wt x_2^2+\wt x_3^2} & \frac{\wt x_3}{-\alpha
   ^2+\wt x_2^2+\wt x_3^2} & -\frac{\wt x_2}{-\alpha
   ^2+\wt x_2^2+\wt x_3^2} & 0 \\
 -\frac{\wt x_3}{-\alpha ^2+\wt x_2^2+\wt x_3^2} &
   \frac{\wt x_2^2-\alpha ^2}{-\alpha ^2+\wt x_2^2+\wt x_3^2} &
   \frac{\wt x_2 \wt x_3}{-\alpha ^2+\wt x_2^2+\wt x_3^2} & 0 \\
 \frac{\wt x_2}{-\alpha ^2+\wt x_2^2+\wt x_3^2} & \frac{\wt x_2
   \wt x_3}{-\alpha ^2+\wt x_2^2+\wt x_3^2} & \frac{\wt x_3^2-\alpha
   ^2}{-\alpha ^2+\wt x_2^2+\wt x_3^2} & 0 \\
 0 & 0 & 0 & 1
\end{array}
\right)
\end{equation*}gives a metric with nonvanishing scalar curvature
\begin{equation*}
    \wwt R=-\frac{10 \alpha ^2+4
   \left(\wt x_2^2+\wt x_3^2\right)}{\left(\wt x_2^2+\wt x_3^2-\alpha
   ^2\right)^2},
\end{equation*}so it cannot be transformed to the pp-wave form. On the other hand, the
metric of this background can be diagonalized to the form
\begin{equation}\label{dualmtkBrink18}
ds^{2}= \frac{1}{y_3^2-\alpha ^2}\,d{y_1}^2+\frac{y_3^2 \alpha ^2}{\alpha
^2-y_3^2}\,d{y_2}^2+d{y_3}^2+d{y_4}^2,
\end{equation} by
\begin{equation*}\label{BrinkCoors18}\wt
x_1=y_1,\quad \wt x_2=y_3\cos y_2,\quad \wt x_3=y_3\sin y_2,\quad \wt x_4=y_4.
\end{equation*}
Note the singularity on the surfaces $y_3=\pm \alpha $. For $|y_3|<\alpha$ the time-like
direction is given by the vector $\partial_{y_1}$, whereas for $|y_3|>\alpha$ the time-like
vector  is $\partial_{y_2}$.

The torsion acquires the form
$$ H=\frac{2 y_3 \alpha ^2}{\left(y_3^2-\alpha
   ^2\right)^2}\,dy_1\wedge dy_2\wedge dy_3$$
and the dilaton satisfying \eqref{bt1}--\eqref{bt3} is}
$$ \Phi=\log(y_3^2-\alpha^2) + const.$$

{To find the solution of the \eqn s of this dual \sm {}, we need the above \tfn{} between
$y_j$ and $\wt x_j$, and also $\wt x_j$ expressed in terms of $x^j,\wt h_k$. As the
commutation relations are {the same as for $S_{17}$ and $S_{19}$,} we get}
\begin{align*}
    \wt x_1 &= \wt h_1+x^2 \wt h_3-x^3 \wt h_2,
   & \wt x_2 &=\wt h_2 \cos x^1-\wt h_3\sin x^1,\\
   \wt x_4 &= \wt h_4,
   & \wt x_3 &= \wt h_2\sin x^1+\wt h_3\cos x^1.
\end{align*}
\section{Conclusion}\label{conclusion}
We have classified all atomic non-Abelian duals of the four-dimensional flat spacetime with
respect to four-dimensional subgroups of the Poincaré group. As a result, we have obtained 14
different types of exactly solvable \sm s in the four-dimensional curved
backgrounds. %obtained as atomic non-Abelian T-duals of the flat metric.
Due to the non-Abelian T-duality, one can find {general solutions of the classical field equations for all of these dual models in terms of d'Alembert solutions of the wave \eqn{}}. The method of obtaining the solutions is described {in section \ref{solvePLT}} {and examples are given in section \ref{examples} {and \ref{results}}.}
{One-loop beta equations for all of the dual backgrounds yield simple ordinary differential equations for dilatons. Their solutions are given in sections \ref{examples} {and \ref{results}}.}

Eleven of the dual backgrounds are plane-parallel waves whose metrics can be brought to the
Brinkmann form
\begin{equation*}\label{BrinkMetrics}
   ds^{2}= 2dudv-[K_3(u){z_3}^2+K_4(u){z_4}^2]du^2+d{z_3}^2+d{z_4}^2.
\end{equation*}
The torsion then is
\begin{equation*}\label{BrinkTorsion}
   H=dB=H(u)\,du\wedge dz_3\wedge dz_4.
\end{equation*}
{Depending on the chosen subgroup,} functions $K_3(u),K_4(u),H(u)$ acquire various forms, as follows:
\begin{eqnarray}
\label{homogeneous model}   K_3(u) =K_4(u)=1, & H(u)=-2, \\
\label{mtk 23 25}   K_3(u) = \frac{3} {\left(u^2+1\right)^2},\ \
K_4(u)=-\frac{\left(2 u^2-1\right)} {\left(u^2+1\right)^2}, &  H(u)=\pm\frac{2}{u^2+1},\ \ \ \  \\
\label{mtkA 7 8 26 27}   K_3(u) =2\,\text{sech}^2(u),
\ \ K_4(u)=2\,\delta\,\text{sech}^2(u),\ \delta=0,1,&H(u)=0, \\
\nonumber  \\
\label{mtkB 7 8 26 27}   K_3(u) =-2\,\text{csch}^2(u),
\ \ K_4(u)=-2\,\delta\,\text{csch}^2(u),\ \delta=0,1,&H(u)=0, \\
\nonumber   \\
\label{mtk 28A}   K_3(u) =K_4(u)= \frac{
 \left(1+2 \beta ^2 \text{sech}^2({u})\right) }{ \beta ^2}, &H(u)=-\frac{2}{\beta},\\
\label{mtk 28B}   K_3(u) =K_4(u)= \frac{
 \left(1-2 \beta ^2 \text{csch}^2({u})\right) }{ \beta ^2}, &H(u)=-\frac{2}{\beta},
\end{eqnarray}

\begin{eqnarray}
\nonumber  K_3(u) &=&-\frac{\text{sech}^4(u) \left(2 \left(\beta ^2+1\right)
\sinh ^2(u)-\beta ^2\right)}{\left(\tanh ^2(u)+\beta ^2\right)^2},\\
\label{mtk 31 33 tanh}  K_4(u) &=&\frac{\beta ^2 \text{sech}^4({u}) \left( 2 (\beta ^2+1)
\cosh^2({u})+1\right)}{\left(\tanh^2({u})+\beta ^2\right)^2},
\\ \nonumber H(u)&=& \frac{2 \beta}{\beta ^2\cosh^2({u})+\sinh^2({u})},
\end{eqnarray}

\begin{eqnarray}
\nonumber K_3(u) &=&\frac{\text{csch}^4(u) \left(2(\beta ^2+1)\cosh ^2(u)+\beta ^2 \right)}{\left(\coth ^2(u)+\beta ^2\right)^2} ,\\
\label{mtk 31 33 coth} K_4(u) &=& -\frac{\beta ^2 \text{csch}^4(u) \left(2(\beta ^2+1)\sinh ^2(u)-1\right)}{\left(\coth ^2(u)+\beta ^2\right)^2},
   \\ \nonumber H(u)&=& -\frac{2 \beta}{\cosh^2({u})+\beta ^2\sinh^2({u})},
\end{eqnarray}where $\beta\in\real\smallsetminus\{0\}.$

Even though the B-fields obtained by T-duality are usually not of the form
$B=B_i(u)\,du\wedge dz_i$,
%torsions are of the form $H=H(u)\,du\wedge dz_3\wedge dz_4$, so that the B-fields
they are gauge equivalent to $$B'=H(u)\,du\wedge( z_3dz_4-z_4dz_3),$$ and the corresponding \sm
s are exactly conformal \cite{BLPT}. { %All of these pp-wave backgrounds but

Except for (\ref{mtk 28A}), (\ref{mtk 28B}), {these pp-wave backgrounds can be
transformed to the gauged WZW background forms \eqref{sfets} by the standard transformation
from Brinkmann to Rosen coordinates \cite{blaulough}. In most of the transformed
backgrounds the function $g_1$ acquires the form $g_1(u)=1$ and the function $g_2$ acquires the form of one of the functions \eqref{g1g2}, but some other combinations of functions $(g_1, g_2)$ also arise,
namely $(u^{-2},\tanh^2u)$, $(u^{-2},\coth^2u)$, $(\tanh^2u,\tanh^2u)$ and
$(\coth^2u,\coth^2u)$.

Consequently, the pp-waves of the form \eqref{sfets} are duals of the flat metric not only
for $g_1(u)=1$ and $g_2(u)=u^2$, as mentioned in section \ref{intro}, but also for many other
combinations of functions $g_1, g_2$ from the set \eqref{g1g2}.}

It is a remarkable fact that {duals with respect to subgroups corresponding to} non-isomorphic
algebras may lead to the same backgrounds (up to coordinate \tfn{}). These are the cases of
metric \eqref{homogeneous model} produced by subalgebras $S_{17}$ and $S_{29}$,  and also
metrics \eqref{mtkA 7 8 26 27}, \eqref{mtkB 7 8 26 27} obtained from $S_{7}$ and $S_{8}$
for $\delta=0$, and $S_{26}$ and $S_{27}$ for $\delta=1$. The metric \eqref{homogeneous
model} is apparently a homogeneous exactly solvable model with nontrivial constant torsion.
On the other hand, isomorphic (but not equivalent under proper ortochronous Poincaré
transformations) algebras $S_{23}$ and $S_{25}$ give the same metric, namely \eqref{mtk 23 25}, but
opposite torsions. Isomorphic algebras $S_{31}$ and $S_{33}$ give the same metrics and
torsions \eqref{mtk 31 33 tanh}, \eqref{mtk 31 33 coth}.

{We also get, besides the pp-waves, dual metrics with nonvanishing scalar curvature and torsion:
\begin{equation}\label{dualmtkBrink11b}
ds^{2}= -d{y_1}^2+d{y_2}^2+\frac{y_1^2 }{y_1^2+\alpha ^2}\,d{y_3}^2+\frac{1}{y_1^2+\alpha
^2}\,d{y_4}^2,
\end{equation}
$$ H=-\frac{2 y_1 \alpha }{\left(y_1^2+\alpha
   ^2\right)^2}\,dy_1\wedge dy_3\wedge dy_4,$$
\vskip .5cm
\begin{equation}\label{dualmtkBrink18b}
ds^{2}= \frac{1}{y_3^2-\alpha ^2}\,d{y_1}^2+\frac{y_3^2  }{\alpha
^2-y_3^2}\,d{y_2}^2+d{y_3}^2+d{y_4}^2,
\end{equation}$$ H=\frac{2 y_3 \alpha}{\left(y_3^2-\alpha
   ^2\right)^2}\,dy_1\wedge dy_2\wedge dy_3$$
\vskip .5cm
\begin{equation}\label{dualmtkBrink19b}
ds^{2}= -d{y_1}^2+d{y_2}^2+\frac{y_2^2}{y_2^2+\alpha ^2}\,d{y_3}^2+\frac{1}{y_2^2+\alpha
^2}\,d{y_4}^2
\end{equation}$$
H=\frac{2 y_2 \alpha}{\left(y_2^2+\alpha
   ^2\right)^2}\,dy_2\wedge dy_3\wedge dy_4
$$
%that can be brought to the diagonal forms \eqref{dualmtkBrink19}, \eqref{dualmtkBrink11}, \eqref{dualmtkBrink18}.
They are  obtained as non-Abelian duals with respect to $S_{11}, S_{18}, S_{19}$. Note that
isomorphic (but not equivalent under proper ortochronous Poincaré transformation)
subalgebras $S_{17}$ (resp., $S_{18}, S_{19}$) lead to backgrounds
%that differ dramatically
with vanishing (resp., nonvanishing) curvature.

The metrics \eqref{dualmtkBrink11b}--\eqref{dualmtkBrink19b} remind us of black hole
\cite{horava:blackhole} and cosmological backgrounds \cite{nappiw92} rewritten in
\cite{tsey:revExSol} into {diagonal forms depending again on particular  functions $g_1,
g_2$.
The difference from \eqref{dualmtkBrink11b}--\eqref{dualmtkBrink19b} %from those backgrounds
lies in these functions.}
%%%%%%%%%%%%%%%%%%%%%%%%%%%%%%%%%%%%%%%%%%%%%%%%%%%%%%%%%%%%%%%%%%%%%%%%%%%%%%%
%% Appendix
%%%%%%%%%%%%%%%%%%%%%%%%%%%%%%%%%%%%%%%%%%%%%%%%%%%%%%%%%%%%%%%%%%%%%%%%%%%%%%%

%\appendix

\section*{Appendix: Poincaré subalgebras}\label{appA}
{We summarize the four-dimensional Poincaré subalgebras that act
freely and transitively on the flat manifold. The numbering of the
subalgebras follows from the order introduced in \cite{PWZ}, Table
IV.}
\begin{align*}
%    S &= Span[\mathcal{K}_1,\mathcal{K}_2,\mathcal{K}_3,\mathcal{K}_4]\\ \\
  S_1 &= Span[P_0, P_1, P_2, P_3] ,&\\
  S_2 &= Span[M_3, P_0 - P_3, P_1, P_2] ,&\\
  S_6 &= Span[L_2 + M_1-\half (P_0+P_3),P_1, P_0 - P_3,P_2] ,&\\
  S_7 &= Span[2  M_3 + \alpha\,P_1, L_2 + M_1,  P_0 - P_3, P_2] ,& \alpha>0,\\
  S_8 &= Span[ M_3 , L_2 + M_1,  P_0 - P_3, P_2] ,&\\
  S_{11} &= Span[ M_3 + \alpha\,P_2,P_0,  P_3, P_1] ,& \alpha>0,\\
  S_{17} &= Span[ L_3+ \epsilon\,(P_0+ P_3), P_1, P_2, (P_0- P_3)] ,& \epsilon=\pm 1\\
  S_{18} &= Span[ L_3 + \alpha\,P_0, P_1, P_2, P_3] ,& \alpha>0,\\
  S_{19} &= Span[ L_3 + \alpha\,P_3, P_1, P_2, P_0] ,& \alpha\neq 0,\\
  S_{23} &= Span[ L_2 + M_1-\half (P_0+P_3),L_1-M_2+\alpha\,P_1, P_0 - P_3,P_2] ,&  \alpha\neq 0,\\
  S_{25} &= Span[ L_2 + M_1-\epsilon P_2, P_0 + P_3,P_1, P_0 - P_3],& \epsilon=\pm 1\\
  S_{26} &= Span[M_3 + \alpha\,P_1, L_2 + M_1, L_1 - M_2, P_0 - P_3] ,&  \alpha>0,\\
  S_{27} &= Span[M_3,              L_2 + M_1, L_1 - M_2, P_0 - P_3] ,&\\
  S_{28} &= Span[L_3 - \beta\,M_3, L_2 + M_1, L_1 - M_2, P_0 - P_3] ,&  \beta\neq 0,\\
  S_{29} &= Span[L_3 - \beta\,M_3, P_0 - P_3, P_1, P_2] ,&  \beta\neq 0,\\
  S_{31} &= Span[M_3,              P_1 + \beta\,P_2, P_0 - P_3, L_2 + M_1] ,&  \beta\neq 0,\\
  S_{33} &= Span[M_3 + \alpha\,P_2, P_1 + \beta\,P_2, P_0 - P_3, L_2 + M_1] ,&  \alpha>0,\beta\neq 0.
\end{align*}

%%%%%%%%%%%%%%%%%%%%%%%%%%%%%%%%%%%%%%%%%%%%%%%%%%%%%%%%%%%%%%%%%%%%%%%%%%%%%%%
%% Backmatter
%%%%%%%%%%%%%%%%%%%%%%%%%%%%%%%%%%%%%%%%%%%%%%%%%%%%%%%%%%%%%%%%%%%%%%%%%%%%%%%

%\backmatter


\begin{thebibliography}{999}
\bibitem{papa}{G. Papadopoulos, J. G. Russo and A. A. Tseytlin,
\emph{Solvable model of strings in a time-dependent plane-wave background}, Class. Quant.
Grav. 20 (2003) 969, [hep-th/0211289].}

\bibitem{tsey:revExSol}{A. A. Tseytlin, {\emph {Exact solutions of closed string theory}},
Class. Quant. Grav. 12 (1995) 2365, [hep-th/9505052].}

\bibitem{BLPT}{M. Blau, M. O'Loughlin, G. Papadopoulos and A.  A. Tseytlin,
 \emph{Solvable models of strings in homogeneous plane wave backgrounds},
 Nucl. Phys. B673 (2003) 57, [hep-th/0304198].}

\bibitem{HorSteif}{G. T. Horowitz and A. R. Steif},
{\emph{Strings in strong gravitational fields}, Phys. Rev. D 42 (1990), 1950 - 1959}.

\bibitem{VegaSan} H. J. de Vega and N. Sanchez,
{\emph {Strings Falling Into Space-Time Singularities}}, Phys. Rev. D
45 (1992) 2783.

\bibitem{Penrose} R. Penrose, {\it Any Space-Time has a Plane Wave as a Limit}, in: Differential Geometry and Relativity, vol. 3, eds. M. Cahen and M. Flato, Reidel, Dordrecht 1976.

\bibitem{Gueven} R. Gueven, {\it Plane wave limits and T-duality}, Phys. Lett. B 482 (2000), 255 - 263, [hep-th/0005061v1].

\bibitem {Sfetseyt94} K. Sfetsos, A. A. Tseytlin,
{\it  Four Dimensional Plane Wave String Solutions with Coset CFT Description}, Nucl. Phys.
B 427 (1994) 245, [hep-th/9404063].

\bibitem{tseyt94} A. A. Tseytlin, {\it Exact string solutions and duality}, [hep-th/9407099].

\bibitem{horava:blackhole} P. Ho\v rava
{\it Some Exact Solutions of String Theory in Four and Five Dimensions}, Phys. Lett. B 278
(1992) 101, [hep-th/9110067].

\bibitem{nappiw92} C. Nappi and E. Witten, \emph{A closed expanding universe in string theory},
Phys. Lett.  B293 (1992) 309,  [hep-th/9206078].

\bibitem{delaossa:1992vc}
X. C. de~la Ossa and F.~Quevedo, {\it Duality symmetries from non-abelian
  isometries in string theories},  Nucl. Phys. {B403} (1993) 377,
  [hep-th/9210021].

\bibitem{klise}{C. Klim\v c\'ik and P. \v Severa, {\emph{Dual non-Abelian duality and the Drinfeld double}},
Phys. Lett. B 351 (1995) 455, [hep-th/9502122].}

\bibitem{dhh:vca}{C. Duval, Z. Horv\'ath, and  P. A. Horvathy,  {\emph {Vanishing of the conformal anomaly for strings
in a gravitational wave}}, Phys. Lett. {B313} (1993) 10, [hep-th/0306059].}

\bibitem{dhh:spfgw}{C. Duval, Z. Horv\'ath, and  P. A. Horvathy,
{\emph {Strings in plane-fronted gravitational waves}},  Mod. Phys. Lett. {A8} (1993) 39,
[hep-th/0602128].}

\bibitem{klim:proc}{C. Klim\v c\'ik, \emph{Poisson--Lie T-duality}, Nucl. Phys. B
(Proc. Suppl.) 46 (1996) 116, [hep-th/9509095].}

\bibitem{PWZ} J. Patera, R. T. Sharp, P. Winternitz and H. Zassenhauss,
{\emph {Subgroups of the Poincaré group and their invariants} }, J. Math. Phys. 17 (1976)
977.

\bibitem{hlapet:ijmp} L. Hlavat\'y and I. Petr, \emph{New solvable  sigma models in plane-parallel wave
background}, Int. J. Mod. Phys. A29 (2014) 1450009, [arXiv:1308.0153]

\bibitem{blaulough}{M. Blau and M. O'Loughlin,
\emph{Homogeneous Plane Waves}, Nucl. Phys. B 654 (2003) 135--176, [hep-th/0212135].}

\end{thebibliography}
\end{document}